\documentclass[aps,prb,preprint]{revtex4-2}

\usepackage{graphicx}
\usepackage{hyperref}
\usepackage{amsmath}
\usepackage{color}

\begin{document}


\title{Correction of Density-Functional-Theory based polynomial interatomic potentials to reproduce experimental melting properties}

\renewcommand{\figurename}{FIG.}
\renewcommand{\tablename}{TAB.}



\author{Bernd Bauerhenne}
\email{bauerhenne@uni-kassel.de}
\author{Martin E. Garcia}
\affiliation{Theoretical Physics and Center for Interdisciplinary Nanostructure Science and Technology (CINSaT), University of Kassel, Heinrich-Plett-Strasse 40, 34132 Kassel, Germany}


\date{\today}

\begin{abstract}
Recently, we developed a method to construct polynomial interatomic potentials from ab-initio calculations in order to accurately describe laser excited solids [PRL 124, 085501 (2020)]. 
However, ab-initio methods, and therefore analytical potentials derived from them, commonly do not provide an accurate prediction of the melting temperature. 
In order to reproduce the experimental melting properties, but keeping the accuracy in the laser excited case, we present here an approach to modify  few key coefficients of polynomial interatomic potentials constructed from ab-initio data. 
We show that, with the help of such corrections, the electronic-temperature dependent interatomic potential for silicon can, at the same time, describe nonthermal laser induced effects with ab-initio accuracy and also provide the correct experimental melting temperature and slope $dT/dp$.
\end{abstract}


\maketitle

\section{Introduction}

Interatomic potentials allow for ultra-large scale atomistic molecular dynamics (MD) simulations with up to billions of atoms \cite{Shibuta2017} and simulation times of nanoseconds, which is necessary to get insights into many physical processes, such as diffusion \cite{Cheng2018,Sushko2014,Hoy2000}, plastic deformation \cite{Verkhovtsev2013,Zink2006}, melting \cite{Cleveland1998,Qi2001}, crystallization \cite{Qi2001,Yakubovich2013} and other phase transformations \cite{Kexel2015,Pun2010}.
Femtosecond laser pules excite the electrons in matter to high electronic temperatures $T_\text{e}$'s inducing significant ultrafast changes in the interatomic bonding whereas the ions remain mostly unaffected until electron-phonon interactions become active \cite{Stampfli1990}.
In order to address the short lived changes in interatomic bonding due to the hot electrons in large scale MD simulations, $T_\text{e}$-dependent interatomic potentials were introduced \cite{Khakshouri2008,Murphy2015,Norman2012,Moriarty2012,Shokeen2010,Shokeen2011,Darkins2018,Bauerhenne2020}, which depend beside the atomic coordinates also on the electronic temperature $T_\text{e}$.
The hot electrons cause many ultrafast phenomena like bond hardening or softening \cite{Recoules2006,Grigoryan2014,Fritz2007}, structural solid-solid and solid-liquid phase transitions \cite{Cavalleri2001,Sciaini2009,Buzzi2018}, phonon squeezing or antisqueezing \cite{Johnson2009,Zijlstra2013Jan}, excitation of coherent phonons \cite{Cheng1991,Hase2003}, which can be well described by  $T_\text{e}$-dependent density functional theory (DFT).
Such ab-initio methods cannot access atomistic simulations on such large temporal and spatial dimensions as interatomic potentials can do.
It has been shown in different works that one can use ab-initio methods to generate data for constructing interatomic potentials \cite{Tersoff1986,Stillinger1985}.
However, an accurate prediction of the melting properties of solids is usually not feasible using ab-initio methods.
In silicon (Si), for instance, DFT in the local density approximation (LDA) predicts a melting temperature of $T_\text{m}(p)= (1300 \pm 50)\, \text{K} - 58\, \frac{\text{K}}{\text{GPa}}\times p$ \cite{Alfe2003}, which is 20 \% below the experimental value of $T_\text{m}=(1687 \pm 5)\, \text{K} - 58\, \frac{\text{K}}{\text{GPa}} \times p$ \cite{Yamaguchi2002,Jayaraman1963}.
The usage of the generalized gradient approximation of Perdew-Burke-Ernzerhof (PBE) improves the prediction to $T_\text{m}(p)= (1492 \pm 50)\, \text{K} - 42\, \frac{\text{K}}{\text{GPa}}\times p$ \cite{Alfe2003}.
But only the application of the random phase approximation (RPA) together with PBE yields the correct melting temperature \cite{Dorner2018}.
Although compact analytical expressions exist for the interatomic forces within the RPA, the computation of the forces is extremely  demanding \cite{Ramberger2017}.
Thus, in order to construct an accurate interatomic potential, it may be more efficient to generate the data from DFT without RPA, to fit these data to an interatomic potential and to modify afterwards the coefficients of the obtained interatomic potential for reproducing the experimental melting temperature.
For example, this was done by Kumagai {\it et al.}, who developed an interatomic potential for Si with electrons in the ground state.
The coefficients were firstly fitted to LDA-DFT data and then one coefficient was additionally modified to reach the experimental melting temperature \cite{Kumagai2007}.
Such a procedure may be even more appealing for constructing $T_\text{e}$-dependent interatomic potentials, since $T_\text{e}$-dependent DFT alone describes very well interatomic bonding at increased $T_\text{e}$ and, therefore, is quite suitable to generate data for fitting.

We recently developed a $T_\text{e}$-dependent interatomic potential for Si \cite{Bauerhenne2020} by fitting interatomic forces and structural energies from molecular dynamics simulations in thin-film geometry using $T_\text{e}$-dependent DFT in the local density approximation.
The obtained interatomic potential describes, when included with atomistic simulations,  femtosecond laser-induced effects in Si, like the bond softening, thermal phonon antisqueezing, non-thermal melting, and ablation with remarkable accuracy.
The interatomic potential for Si has a melting temperature of $T_\text{m}(p)= (1199 \pm 2)\, \text{K} - (40\pm 3)\, \frac{\text{K}}{\text{GPa}} \times p$ which agrees with the LDA-DFT value but differs from the experimental one.
Here we present a method to modify several coefficients of the $T_\text{e}$-dependent interatomic potential for Si at low $T_\text{e}$'s in such a way that the experimental melting temperature is reproduced, a negative slope in the melting temperature vs. pressure diagram is obtained, and that there are no significant changes in the description of the potential energy surface at high $T_\text{e}$'s.

The paper is organized as follows. 
At first we explain how we calculate the melting temperature and we describe the functional form of our derived $T_\text{e}$-dependent interatomic potential for Si.
Then, we introduce a modification of the three-body term of the potential that increases the melting temperature to the experimental value.
But this modification causes an unphysical positive slope of the melting temperature as a function of pressure close to zero pressure.
To correct this, we introduce a simple trial interatomic potential consisting of a two-body and a three-body potential like the, for instance, the terms present in the Stillinger \& Weber potential \cite{Stillinger1985}.
We modify independently the strength of the two-body and the three-body potential and analyze the influence on the melting temperature and on the slope in the melting temperature vs. pressure curve.
Using the insights learned from this study, we introduce a modification of the two-body and three-body potential of our $T_\text{e}$-dependent interatomic potential for Si that increases the melting temperature to the experimental value and induces a negative slope.
Finally, we analyze the modified potential and show the influence of the modification on the physical properties compared to the original potential.

\section{Methods}

\subsection{Calculation of the melting temperature}
\label{sec:Tm_calculation}

\begin{figure}[htb!]
	\centering
	\includegraphics[width=0.48\textwidth]{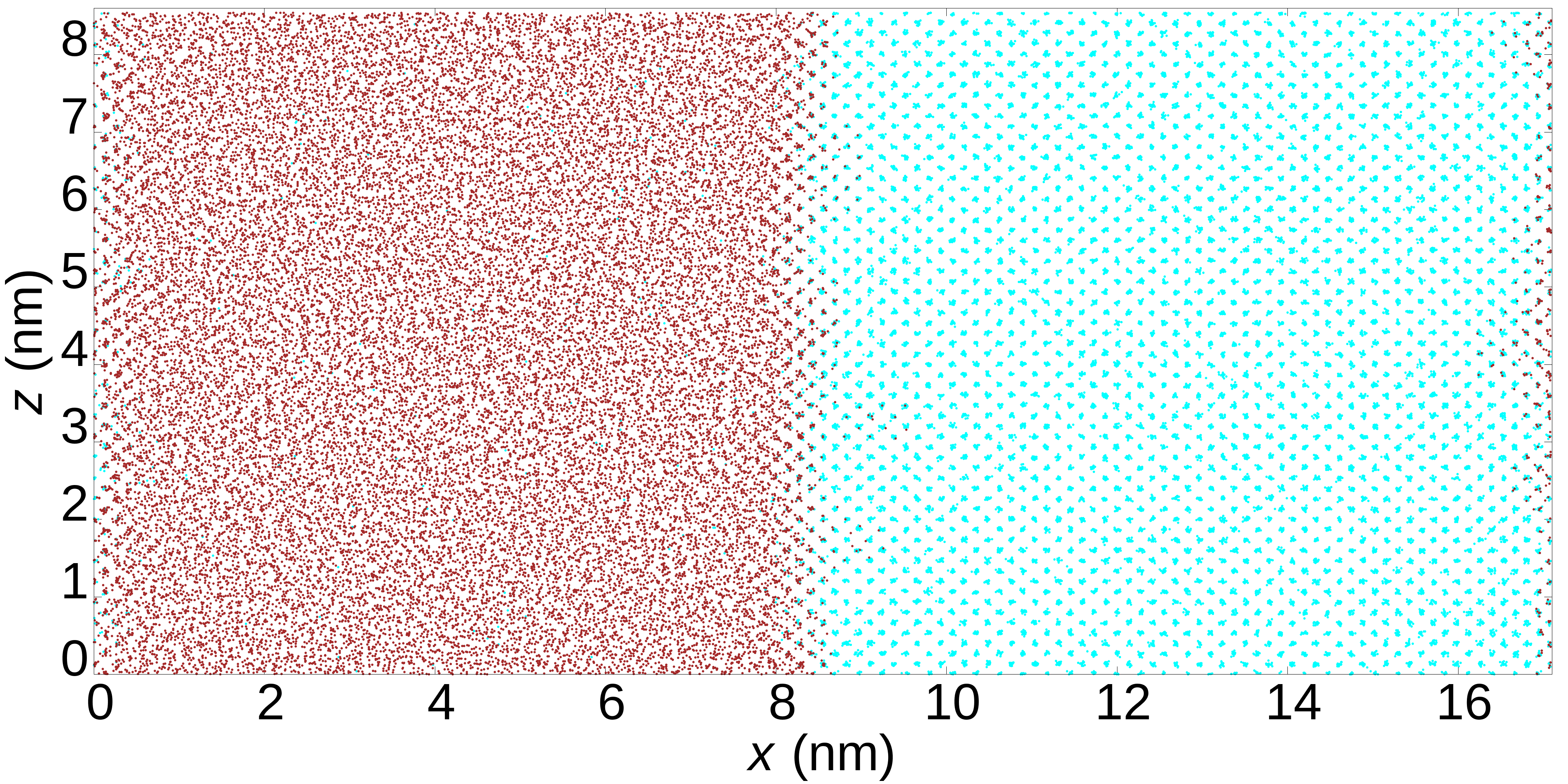}
	\includegraphics[width=0.48\textwidth]{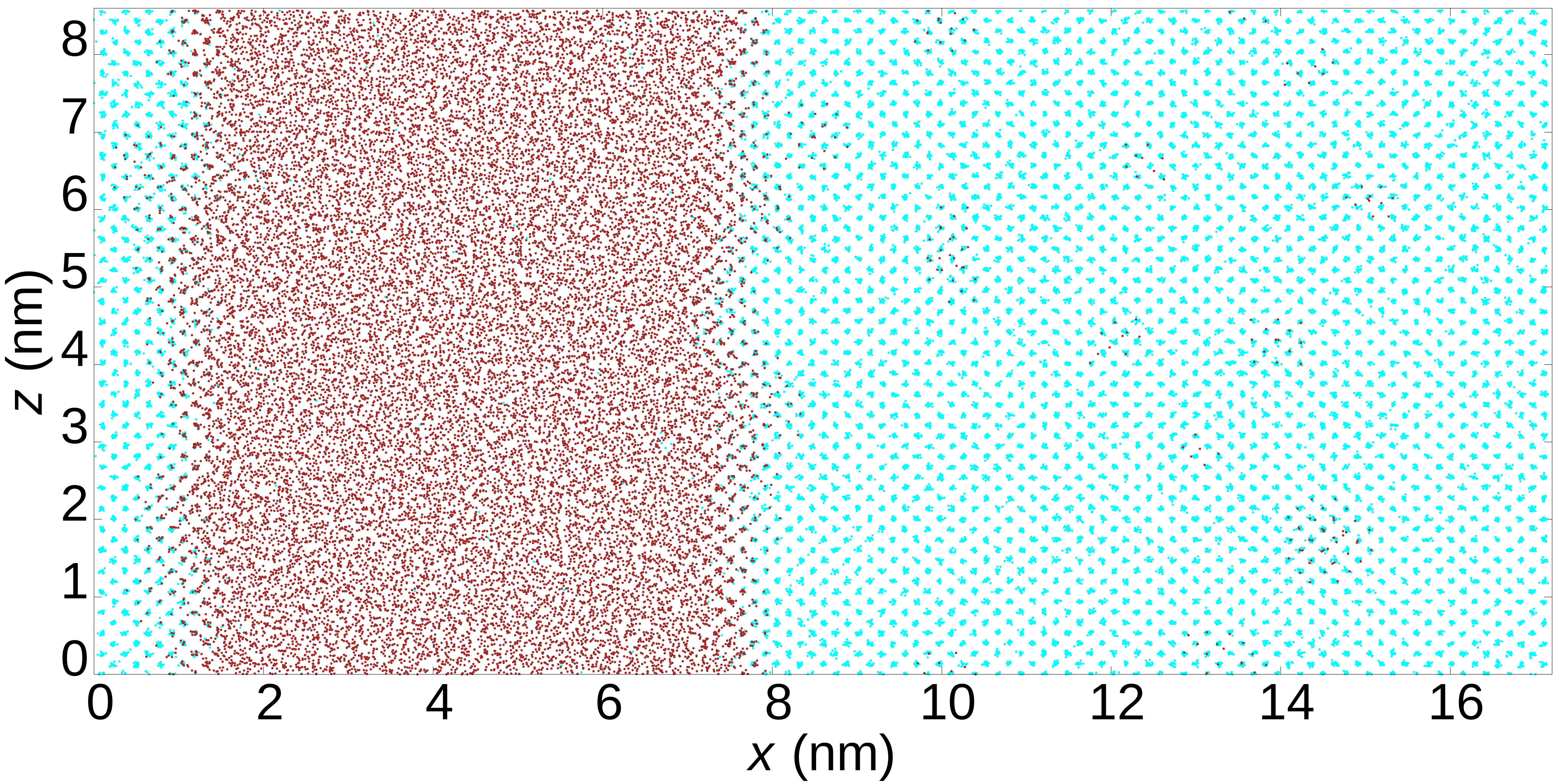}
	\caption{Snapshots of the MD simulation with 65536 Si atoms at zero pressure are shown at the initialization $t=0$ ps (top) and at $t=500$ ps (bottom). 
	The atoms are colored due to their CSP value \cite{Lipp2014}: cyan corresponds to crystalline and brown to molten environment.}
	\label{fig:Si_Polypot1_Si_rxryrz_coexistense}
\end{figure}

We derive the ionic temperature of the ions using the equipartition theorem
\begin{equation}
	T_\text{i} = \frac{2\, E_\text{kin}}{3\, N_\text{at}\, k_\text{B}},
	\label{equ:Ti_definition}
\end{equation}
where $E_\text{kin}$ is the kinetic energy of the ions, $N_\text{at}$ is the number of atoms in the simulation cell and $k_\text{B}$ is the Boltzmann constant.
One cannot derive the melting temperature by just heating up the ideal bulk crystal structure or cooling down the liquid structure.
Thus, we simulated the coexistence of liquid and crystal parts \cite{Keblinski2002} to obtain reliably the melting temperature $T_\text{m}$.
For this, we used a bulk simulation cell that consists of $32\times 16\times 16$ conventional cells and contains $N_\text{at}=65536$ Si atoms.
First we fixed the coordinates of half of the atoms and melted the other part of the crystal structure by applying the Anderson thermostat \cite{Andersen1980} at $T_\text{i}=2500$ K.
Then we allowed again a movement of all atoms.
We applied the Anderson thermostat to all atoms at a temperature assumed to be close to the melting temperature.
During this thermalization, the simulation cell volume and the atomic coordinates were scaled every picosecond to reach a given target pressure.
We did three of such thermalizations in order to get the pressures $p = -1$ GPa, $0$ GPa, $1$ GPa.
Using the above mentioned procedure, we obtained an initialization of atomic coordinates and velocities at a given pressure and temperature.
Half of the structure is molten and the other half is crystalline, so that two planar liquid-crystal interfaces exist within the simulation cell. 
Starting from this initialization, we performed a MD simulation at constant volume and energy for $500$ ps using the Velocity Verlet algorithm \cite{Swope1982}.
We show a snapshot of the atomic structure at the initialization and one after the MD simulation period of $500$ ps for the initialization at zero pressure in Fig. \ref{fig:Si_Polypot1_Si_rxryrz_coexistense}.

\begin{figure}[htb!]
	\includegraphics[width=0.48\textwidth]{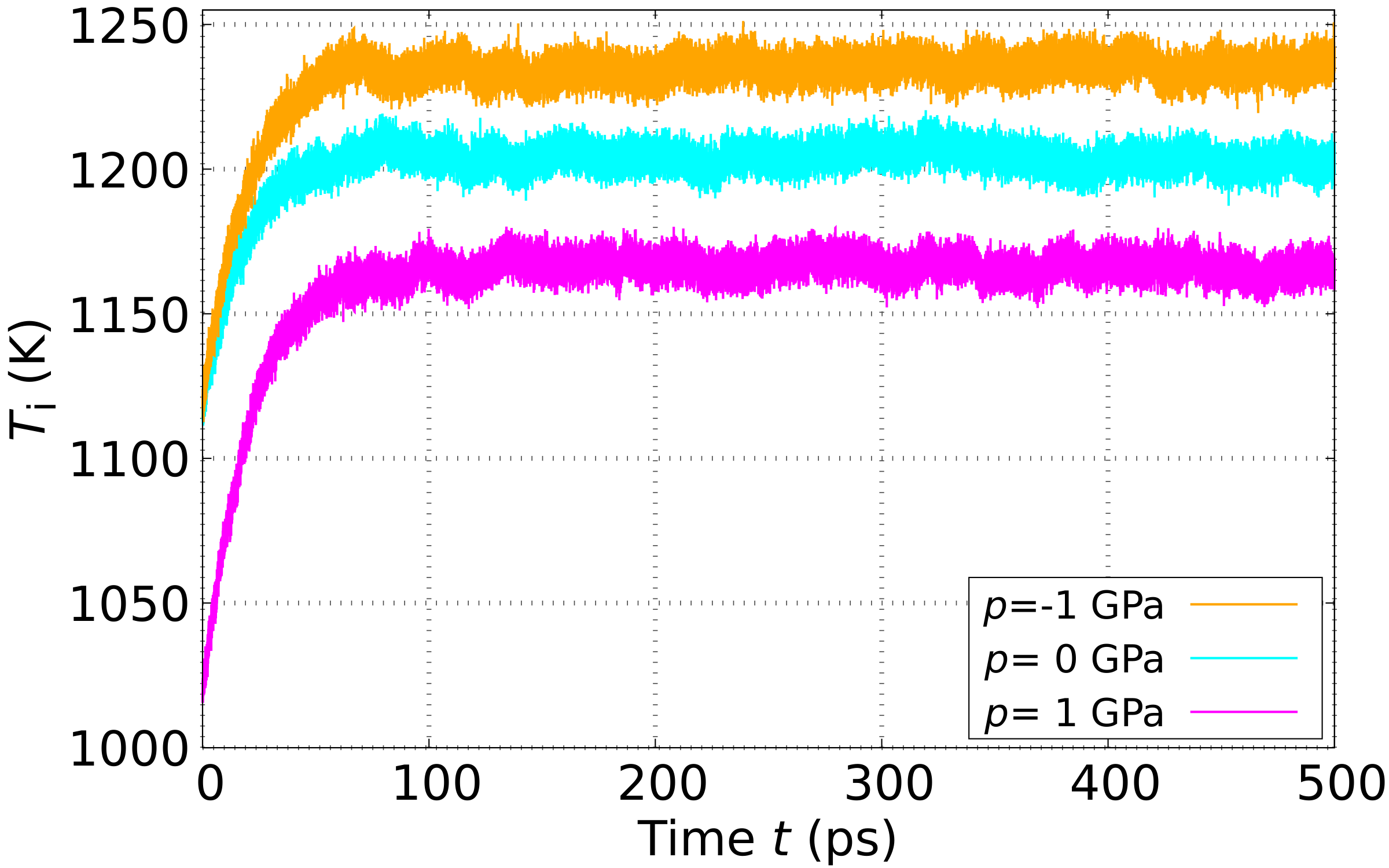}
	\caption{Ionic temperatures occurring in the MD simulation of the bulk simulation cell with 65536 Si atoms are shown as a function of time for various constant pressures.}
	\label{fig:Si_Polypot1_bulk_Tm}
\end{figure}

In this MD simulation at constant energy (and volume), the ionic temperature $T_\text{i}$ converges always to the melting temperature $T_\text{m}$.
This can be explained as followed:
If the initial temperature is below the melting temperature, the atoms of the liquid at the interface start to crystallize, so that the size of the liquid part is decreasing.
This crystallization increases the temperature up to the melting temperature, because the heat of fusion is released from the crystallization.
If the initial temperature is above the melting temperature, the atoms of the crystal at the interface start to melt, so that the size of the crystal part is decreasing.
The melting deceases the temperature down to the melting temperature, because the heat of fusion is taken for the melting.
In both cases the temperature converges to the melting temperature.
If this temperature is reached, it remains constant, since the same amount of atoms melt and crystallize.
Only small fluctuations occur, which decrease with the size of the simulation cell.
We present the ionic temperature $T_\text{i}$ obtained from Eq. \eqref{equ:Ti_definition} of the MD simulations at constant energy as a function of time for the original $T_\text{e}$-dependent interatomic Si potential in Fig. \ref{fig:Si_Polypot1_bulk_Tm}.
One can clearly see, that $T_\text{i}$ converges to the melting temperature at the given pressure and oscillates then around this value.
One should be aware that the temperature also remains constant in the end, if the whole structure melts or crystallizes in the MD simulation at constant energy, because the material remains then molten or crystalline.
This occurs if one initializes the temperature to far away from the melting temperature.
Therefore, we checked additionally the atomic structure, if there is really a coexistence of a liquid-crystal interface, as presented in Fig. \ref{fig:Si_Polypot1_Si_rxryrz_coexistense}.

\begin{figure}[htb!]
	\includegraphics[width=0.48\textwidth]{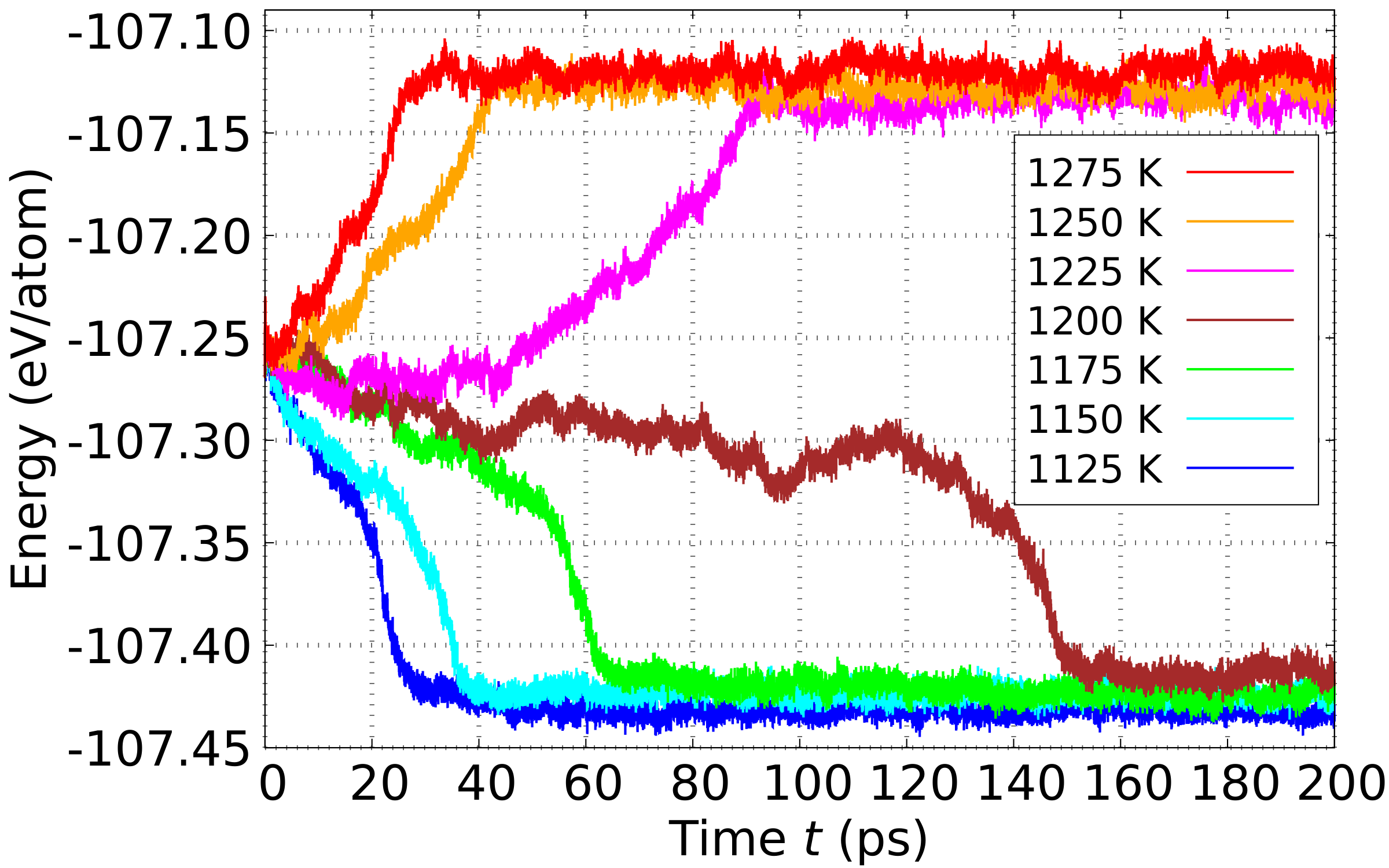}
	\caption{The Helmholtz free energy in the MD simulation of the bulk simulation cell with 1024 Si atoms is shown as a function of time for various temperatures.}
	\label{fig:Si_Polypot1_E_T_smallcells}
\end{figure}

The fluctuations of the temperature, which occur after the melting temperature is reached, decrease with increasing simulation cell size.
On the other hand, if the simulation cell is to small, the fluctuations are so large that the liquid-crystal interface cannot be stabilized, so that the whole structure always melts or crystallizes.
We used this effect occurring in MD simulations of small simulation cells to derive an approximation of the melting temperature.
For this, we set up a simulation cell consisting of $8\times 4\times 4$ conventional cells and containing $N_\text{at}=1024$ Si atoms.
In order to get bulk Si, we used periodic boundary conditions in all directions.
At first, we fixed the atomic coordinates of half of the atoms and applied the Anderson thermostat at $T_\text{i}=2500$ K to the unfixed atoms in order to melt their structure.
By doing this, we obtained a structure, where half is molten and the other half is in a crystalline state.
Then we allow all atoms to move and applied the Andersen thermostat at a given temperature $T_\text{i}$ on a long timescale.
Now the whole structure melts or crystallizes, as one can see in Fig. \ref{fig:Si_Polypot1_E_T_smallcells}, where the structural energy is shown as a function of time for several temperatures.

The energy of the crystal phase is smaller compared to the energy of the liquid phase.
Thus, the structural energy decreases, if the structure crystallizes entirely, and the structural energy increases, if the structure melts entirely.
If the temperature is chosen significantly below the melting temperature, the structure always crystallizes and, if the temperature is chosen significantly above the melting temperature, the structure always melts.
If the temperature is chosen close to the melting temperature, the structure melts or crystallizes depending on the actually used random numbers in the Anderson thermostat.
Due to this, one only obtains a rough approximation of the melting temperature from such simulations. 

Since the melting temperature must be obtained from large-scale MD simulations of the liquid-crystal coexistence, it is obvious that one cannot directly fit the parameters of an interatomic potential to the value of the melting temperature.

\subsection{Analytical form of the interatomic potential}
\label{sec:Polypot}

We construct the $T_\text{e}$-dependent interatomic potential for Si \cite{Bauerhenne2020} as a sum of a two-body potential $\Phi_2$, a three-body potential $\Phi_3$, an embedding function $\Phi_\rho$ and the Helmholtz free energy of an isolated Si atom $\Phi_0$:
\begin{eqnarray}
\Phi &=& \sum_{\scriptsize \begin{array}{c}i<j\\ r_{ij} < r^{(\text{c})}_2 \end{array}} \hspace{-10pt} \Phi_2(T_\text{e},r_{ij}) + \sum_{\scriptsize \begin{array}{c}i\,j\,k\\ r_{ij},r_{ik} < r^{(\text{c})}_3 \end{array}} \hspace{-19pt} ' \ \Phi_3(T_\text{e},r_{ij},r_{ik},\theta_{ijk}) \nonumber \\
& & + \sum_{i} \Phi_{\rho}\left(T_\text{e},\rho^{(2)}_i,\rho^{(3)}_i,\ldots,\rho^{\left(N^{(r)}_\rho\right)}_i\right) \nonumber \\
& & + \sum_{i} \Phi_0(T_\text{e}).
\label{equ:Phi}
\end{eqnarray}
Here $r_{ij}$ denotes the distance between atoms $i$ and $j$, $\theta_{ijk}$ is the angle between $\mathbf{r}_{ij}$ and $\mathbf{r}_{ik}$, the prime indicates that all summation indices are distinct, and $\rho^{(2)}_i$, $\rho^{(3)}_i, \ldots$ are different measures for the atomic density surrounding atom $i$ (see below). 
$r^{(\text{c})}_2$, $r^{(\text{c})}_3$, and $r^{(\text{c})}_\rho$ denote the individual cutoff radii for $\Phi_2$, $\Phi_3$, and $\Phi_\rho$ (see below), respectively.
The different terms are constructed as
\begin{eqnarray}
\Phi_2 &=& \sum_{q=2}^{N_2^{(r)}} c^{(q)}_2\, \left(1-\frac{r_{ij}}{r^{(\text{c)}}_2}\right)^q, 
\label{equ:Phi_2} \\
\Phi_3 &=& \sum_{q_1=2}^{N_3^{(r)}} \ \sum_{q_2=q_1}^{N_3^{(r)}} \ \sum_{q_3=0}^{N_3^{(\theta)}} c^{(q_1\,q_2\,q_3)}_{3} \, \times \nonumber \\
& & \times\left(1-\frac{r_{ij}}{r^{(\text{c})}_3}\right)^{q_1} \left(1-\frac{r_{ik}}{r^{(\text{c})}_3}\right)^{q_2} \Bigl(\cos(\theta_{ijk})\Bigr)^{q_3},
\label{equ:Phi_3} \\
\Phi_\rho &=& \sum_{q_1=2}^{N^{(r)}_\rho} \ \sum_{q_2=1}^{N^{(\rho)}_\rho} c^{(q_1\,q_2)}_{\rho} \, \left(\frac{\rho^{(q_1)}_i}{1+\rho^{(q_1)}_i}\right)^{q_2},
\label{equ:Phi_rho}
\end{eqnarray}
and, for $q_1=2,3,\ldots, N^{(r)}_\rho$, the measures for the atomic density surrounding atom $i$ are constructed as
\begin{equation}
\rho^{(q_1)}_i = \sum_{\scriptsize \begin{array}{c}j\neq i\\ r_{ij} < r^{(\text{c})}_\rho \end{array}} \left(1-\frac{r_{ij}}{r^{(\text{c})}_\rho}\right)^{q_1}.
\label{equ:rho}
\end{equation}
The interatomic potential has the degrees
\begin{equation}
N^{(r)}_2 = 10, \
N^{(r)}_3 = 3, \
N^{(\theta)}_3 = 3 \
N^{(\rho)}_\rho = 2, \
N^{(r)}_\rho = 2
\end{equation}
and needs in total 23 coefficients. 
The two-body potential $\Phi_2$ has 9, the three-body potential $\Phi_3$ has 12, and the embedding function $\Phi_\rho$ has 2 coefficients.
Furthermore, the constant cutoff radii
\begin{equation}
r_2^{(\text{c})}  = 0.63\ {\rm nm}, \quad
r_3^{(\text{c})}  = 0.42\ {\rm nm}, \quad
r_\rho^{(\text{c})} = 0.48\ {\rm nm}
\end{equation}
are used.
The coefficients $\bigl\{c^{(q)}_2\bigr\}$, $\bigl\{c^{(q_1\,q_2\,q_3)}_3\bigr\}$, $\bigl\{c^{(q_1\,q_2)}_\rho\bigr\}$ depend on $T_\text{e}$ and are tabulated in the Supplemental Material of Ref. \cite{Bauerhenne2020}.

\section{Results and Discussion}

\subsection{Correction of the 3-body potential coefficients}
\label{sec:correction_3body}

In order to increase the melting temperature in the interatomic potential description, we have to stabilize the Si crystal, which forms the diamond-like structure.
We mean by a stabilization of a structure that the bonding energy of this structure should become higher within the interatomic potential description. 
Each atom joins four nearest neighbors in the diamond-like structure.
The angle $\theta_{ijk}$ between any of these neighbors is always equal and obeys $\cos(\theta_{ijk})=-\frac{1}{3}$.
Consequently, we should stabilize this angle for the nearest neighbors.
We can easily do this done by adding the following correction term to the three-body potential:
\begin{equation}
\Phi^{(\text{cor})}_3(r_{ij},r_{ik},\theta_{ijk}) = g(r_{ij}) \ g(r_{ik})  \left(\cos(\theta_{ijk})+\frac{1}{3}\right)^2,
\label{equ:phi_cor_3b}
\end{equation}

with $g(r_{ij})\geq 0$ and $g(r_{ik})\geq 0$.
This construction takes care that the preferred nearest neighbor angle of the diamond-like structure is stabilized by the parabola that exhibits its minimum at $-\frac{1}{3}$ for $\cos(\theta_{ijk})$.
The distance function $g(r)$ should be constructed in such a way, that mainly the nearest neighbors are affected, which are located at a distance of 0.234 nm for Si.

In order to add a correction term like Eq. \eqref{equ:phi_cor_3b} to the three-body potential, we use the following construction, which just corresponds to a modification of three existing coefficients:
\begin{eqnarray}
&&\Phi^{(\text{cor})}_3(r_{ij},r_{ik},\theta_{ijk}) \nonumber \\
&=& \underbrace{\sqrt{\aleph_3} \left(1-\frac{r_{ij}}{r^{(\text{c})}_3}\right)^3}_{=g(r_{ij})}  \underbrace{\sqrt{\aleph_3} \left(1-\frac{r_{ik}}{r^{(\text{c})}_3}\right)^3}_{=g(r_{ik})} \left(\cos(\theta_{ijk})+\frac{1}{3}\right)^2 \nonumber \\[5pt]
&=& \hspace{19pt} \aleph_3 \left(1-\frac{r_{ij}}{r^{(\text{c})}_3}\right)^3  \left(1-\frac{r_{ik}}{r^{(\text{c})}_3}\right)^3 \Bigl(\cos(\theta_{ijk})\Bigr)^2  \nonumber \\
&& + \frac{2}{3}\,\aleph_3 \left(1-\frac{r_{ij}}{r^{(\text{c})}_3}\right)^3  \left(1-\frac{r_{ik}}{r^{(\text{c})}_3}\right)^3 \cos(\theta_{ijk})  \nonumber \\
&& + \frac{1}{9}\,\aleph_3 \left(1-\frac{r_{ij}}{r^{(\text{c})}_3}\right)^3  \left(1-\frac{r_{ik}}{r^{(\text{c})}_3}\right)^3 . \label{equ:3b_coeff_corr}
\end{eqnarray}
The strength of the correction is controlled by $\aleph_3$.
$\aleph_3=0$ corresponds to the uncorrected original potential.
We selected the term $\left(1- r / r^{(\text{c})}_3 \right)^3$ instead of $ \left(1- r / r^{(\text{c})}_3 \right)^2$, since power of three converges faster to zero at reaching the cutoff-radius of $r^{(\text{c})}_3=0.42$ nm and, consequently, the correction is more dominated at the nearest neighbor distance of $0.234$ nm.
Adding the above mentioned term to the potential corresponds to add $\frac{1}{9}\, \aleph_3$ to $c^{(3\,3\,0)}_3$, $\frac{2}{3}\, \aleph_3$ to $c^{(3\,3\,1)}_3$, and $\aleph_3$ to $c^{(3\,3\,2)}_3$.
The potential correction should only take place at low electronic temperatures $T_\text{e}$ around the experimental melting temperature $T_\text{m}=1687$ K, since no modifications should be done at higher $T_\text{e}$'s.
Thus, we did the following: 
We obtained the interatomic potential coefficients from a polynomial approximation of the fitted ideal coefficient values at the eleven electronic temperatures of $316$ K ($1$ mHa), $3158$ K ($10$ mHa), $6315$ K ($20$ mHa), $\cdots$, $31578$ K ($100$ mHa).
Consequently, we added the corresponding correction value to the ideal coefficient values for $c^{(3\,3\,0)}_3$, $c^{(3\,3\,1)}_3$, and $c^{(3\,3\,2)}_3$.
Furthermore, we added the correction value at $316$ K and added half of it at $3158$ K, since the correction should only take place at low $T_\text{e}$'s and should smoothly vanish above the experimental melting temperature of $T_\text{m}=1687$ K.
Finally, we approximated the corrected polynomial from these at two low $T_\text{e}$'s shifted ideal coefficient values.

\begin{figure}[htb!]
	\centering
	\includegraphics[width=0.48\textwidth]{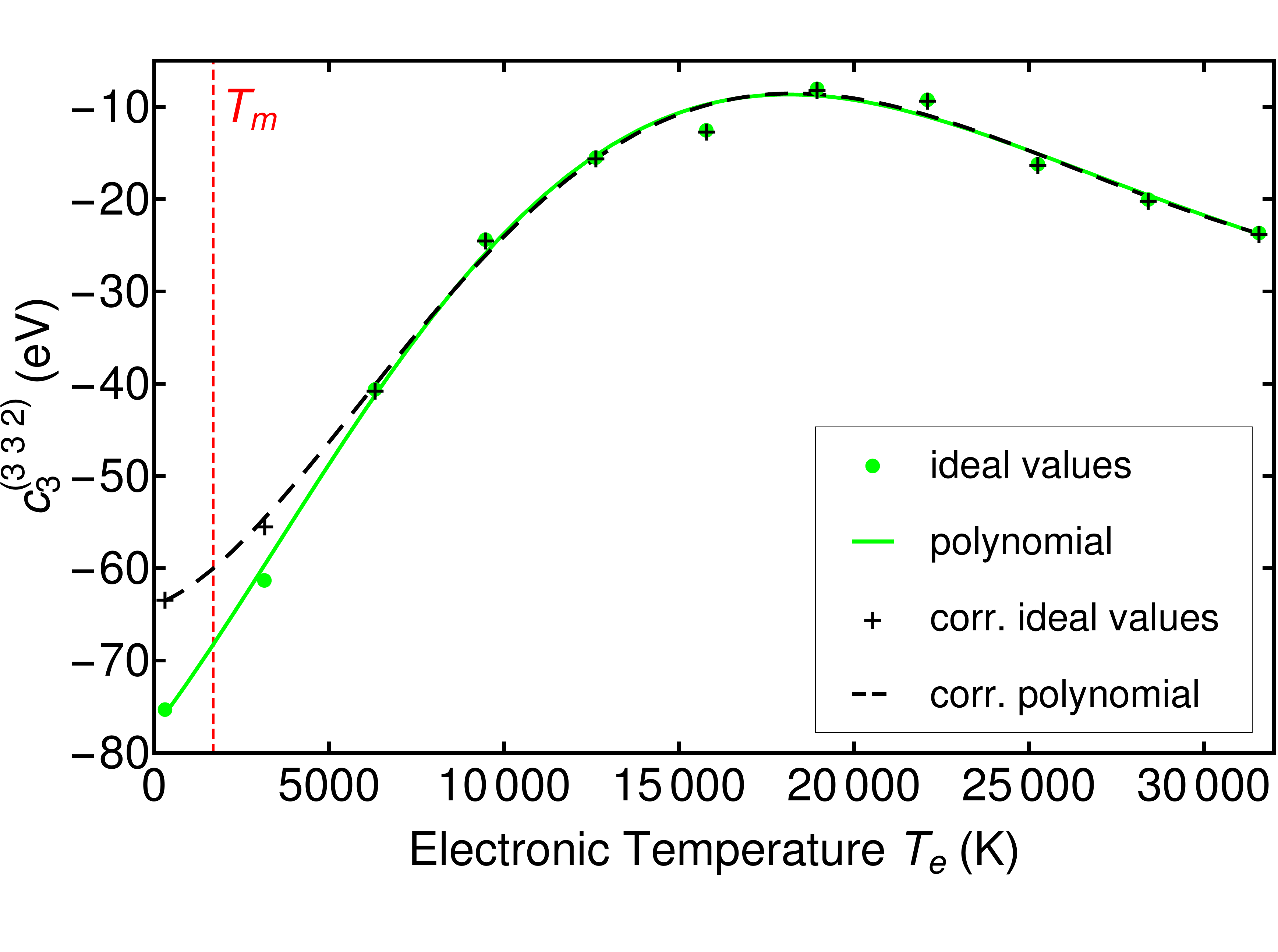}
	\includegraphics[width=0.48\textwidth]{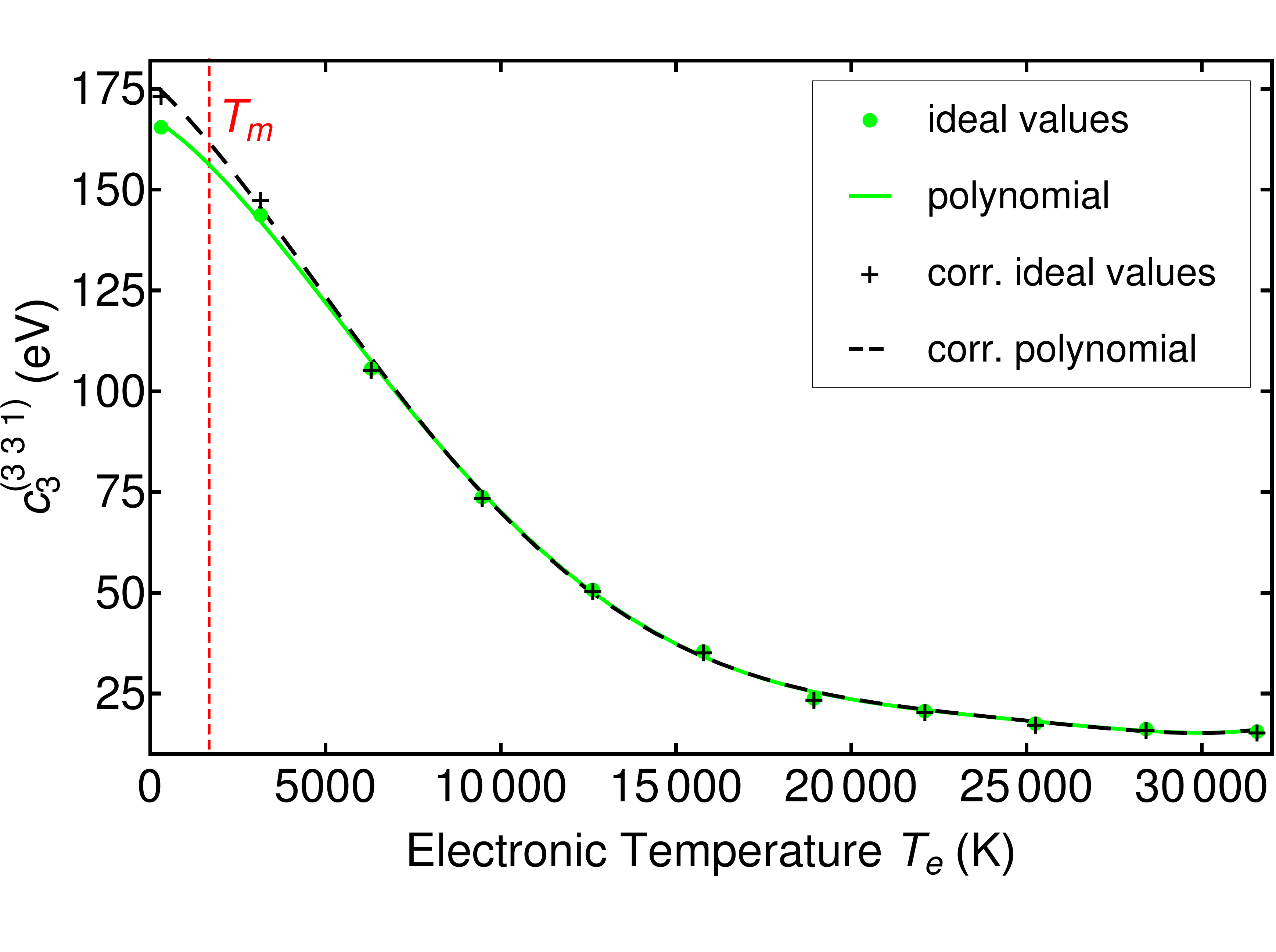}
	\includegraphics[width=0.48\textwidth]{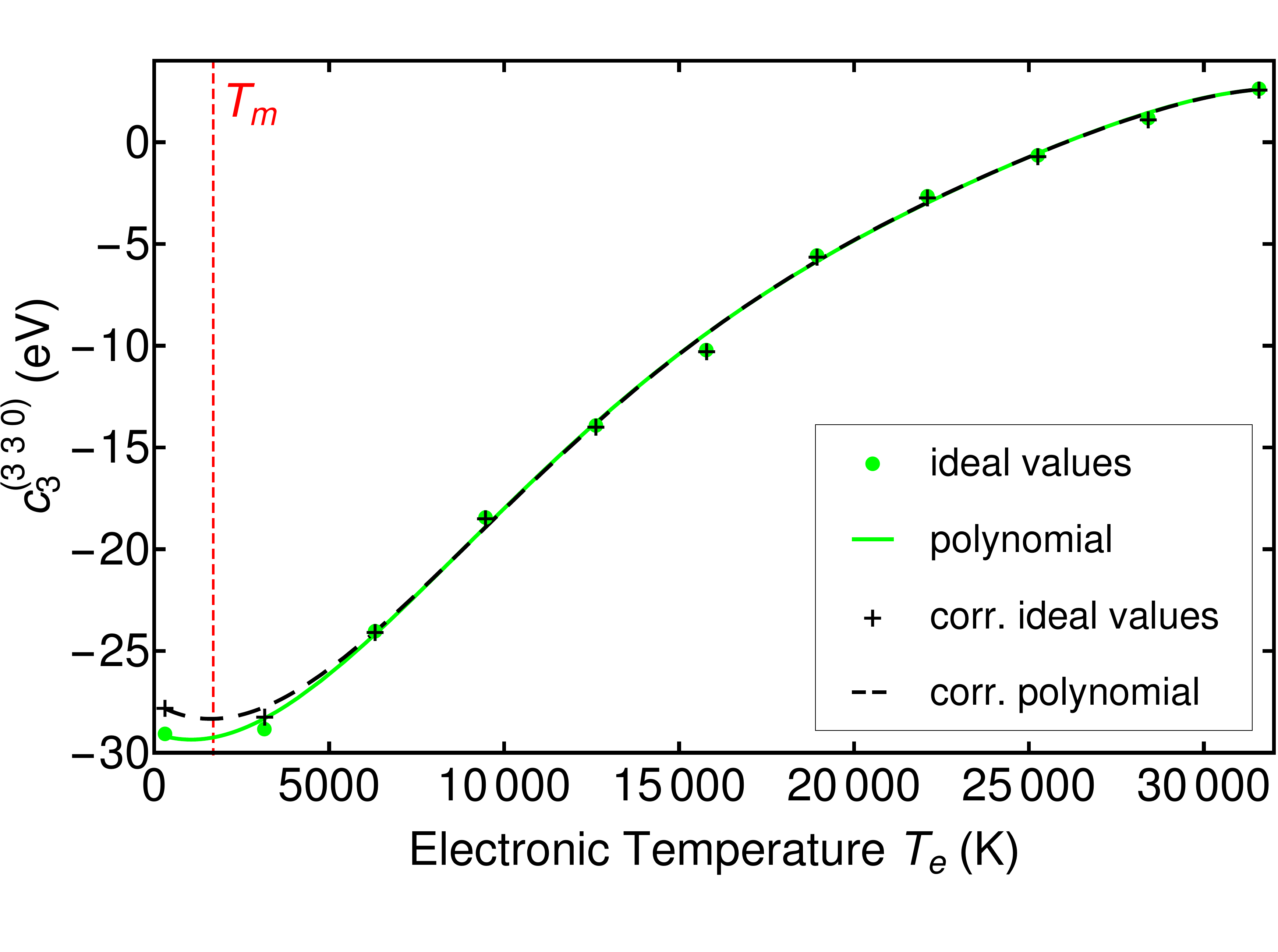}
	\caption{The polynomial approximation of the potential coefficients $c^{(3\,3\,0)}_3$, $c^{(3\,3\,1)}_3$, and $c^{(3\,3\,2)}_3$ is shown together with the ideal values before (green) and after (black) the correction of the ideal values at $T_\text{e}= 316$ K and $T_\text{e}= 3158$ K. $\aleph_3=12$ eV is shown, which yields the experimental melting temperature $T_\text{m}=1687$ K highlighted by a red vertical line.}
	\label{fig:Tm_coeff_corr}
\end{figure}

In order to demonstrate this procedure, we present the original and corrected ideal coefficient values together with the corresponding original and corrected polynomial in Fig. \ref{fig:Tm_coeff_corr} for $\aleph_3=12$ eV, which leads to the experimental melting temperature.
As expected, the corrected polynonial significantly differs from the original only at electronic temperatures below $7000$ K.

We performed the above described correction of the interatomic potential at several values of $\aleph_3$.
We derived the melting temperature $T_\text{m}$ for each corrected interatomic potential at the three pressures $p$ of $-1$ GPa, $0$ GPa, and $1$ GPa.
For this, we performed large scale liquid-crystal coexistence MD simulations using the simulation cell with 65536 atoms as described in Sec. \ref{sec:Tm_calculation}.
We determined the melting temperature $T_\text{m}$ {\it vs.} pressure slope $\bigl.\frac{dT_\text{m}}{dp}\bigr|_{p=0}$ and melting temperature $\bigl.T_\text{m}\bigr|_{p=0}$ at zero pressure from the obtained three $T_\text{m}$ values by a linear regression.
In TAB. \ref{tab:Tm_coeff_correction}, we list the results of the linear regression.

\begin{table}[htb!]
	\caption{\label{tab:Tm_coeff_correction}
		Melting temperature $\bigl.T_\text{m}\bigr|_{p=0}$ and slope $\bigl.\frac{dT_\text{m}}{dp}\bigr|_{p=0}$ in the $T_\text{m}$ {\it vs.} $p$ diagram near zero pressure are listed for different $\aleph_3$-corrections of the interatomic potential coefficients. 
		The listed errors have their origin in the standard deviation of the ionic temperature in the liquid-crystal coexistence MD simulations, because the error of the linear regression is much smaller.}
	\centering
	\begin{tabular}{rlcr}
		\toprule
		$\aleph_3$ (eV) && $\bigl.T_\text{m}\bigr|_{p=0}$ (K) & $\left.\frac{dT_\text{m}}{dp}\right|_{p=0}$ $\left(\frac{\text{K}}{\text{GPa}}\right)$\\
		\hline
		0.0 && $1199 \pm 2$ & $-40 \pm 3$\\
		3.0 && $1388 \pm 2$ & $-12 \pm 3$\\
		6.0 && $1514 \pm 3$ & $  2 \pm 3$\\
		9.0 && $1610 \pm 3$ & $ 11 \pm 4$\\
		12.0 && $1687 \pm 3$ & $ 18 \pm 4$\\
		\hline
	\end{tabular}
\end{table}

An increasing $\aleph_3$ correction induces an increase of the melting temperature of the interatomic potential, as expected, and the experimental value is reached at $\aleph_3=12$ eV.
However, an increasing $\aleph_3$ induces also an increase of the slope $\bigl.\frac{dT_\text{m}}{dp}\bigr|_{p=0}$ in the $T_\text{m}$ {\it vs.} $p$ diagram.
The slope rises from $-40$ K/GPa at $\aleph_3=0$ up to $18$ K/GPa at $\aleph_3=12$ eV (see TAB. \ref{tab:Tm_coeff_correction}).
The experimental value of the slope yields $-58$ K/GPa \cite{Jayaraman1963}.
Consequently, we cannot accept the slope $18$ K/GPa of the melting temperature corrected interatomic potential, because it is even positive compared to the experimental value.

In order to study how an interatomic potential must be modified to increase, on the one hand, the melting temperature and to get, on the other hand, a negative slope, we constructed and analyzed a series of test potentials.
All of these test potentials exhibit the experimental melting temperature, but different slopes. 
The results of this study are reported in the next section.

\subsection{Melting Temperature and slope study on test potentials}
\label{sec:Tm_slope_study_on_test_potentials}

The widely used Stillinger \& Weber potential \cite{Stillinger1985} is the sum of a two-body potential $\Phi^{(\text{SW})}_2$ and a three-body potential $\Phi^{(\text{SW})}_3$.
It exhibits the experimental melting temperature and a significantly negative slope.
The three body potential is constructed like Eq. \eqref{equ:phi_cor_3b} with the distance function 
\begin{equation}
g^{(\text{SW})}(r) = \sqrt{\frac{\lambda_0}{2}}\, \exp\left(\frac{\sigma}{r-r^{(\text{c})}}\right)^\gamma.
\label{equ:SW_g}
\end{equation}
It uses the cutoff radius $r^{(\text{c})}=0.377118$ nm and the parameters $\sigma=0.20951$ nm, $\lambda_0=45.532305023389895$ eV, and $\gamma=1.2$.

In order to construct a simple polynomial test potential $\Phi^{(\text{pol})}$ similar to the construction of the polynomial Si potential, we set it as a sum of a two-body and a three-body potential similar to the Stillinger \& Weber potential.
We construct the two-body potential as
\begin{eqnarray}
\Phi^{(\text{pol})}_2(r_{ij}) &=&  -\frac{3\,\aleph_2 \left(r^{(\text{c})}_2\right)^2}{\left(r^{(\text{c})}_2 - r^{(\text{min})}\right)^2} \left(1 - \frac{r_{ij}}{r^{(\text{c})}_2} \right)^2 \nonumber \\
&& + \quad \frac{2\,\aleph_2  \left(r^{(\text{c})}_2\right)^3}{\left(r^{(\text{c})}_2 - r^{(\text{min})}\right)^3} \left(1 - \frac{r_{ij}}{r^{(\text{c})}_2} \right)^3
\label{equ:Phi2_pol}
\end{eqnarray}
and we construct the three-body potential following Eq. \eqref{equ:phi_cor_3b} using the simple distance function
\begin{equation}
g^{(\text{pol})}(r) = \sqrt{\aleph_3} \left(1 - \frac{r}{r^{(\text{c})}_3} \right)^2.
\label{equ:Phi_pol_g}
\end{equation}

Using this construction, $\Phi^{(\text{pol})}_2$ exhibits one single minimum, which is reached at $r^{(\text{min})}$ and has got the value of $-\aleph_2$.
We set the position of the minimum to $r^{(\text{min})}=0.234$ nm, which is the distance between the nearest neighbors in Si.

\begin{figure}[htb!]
	\centering
	\includegraphics[width=0.48\textwidth]{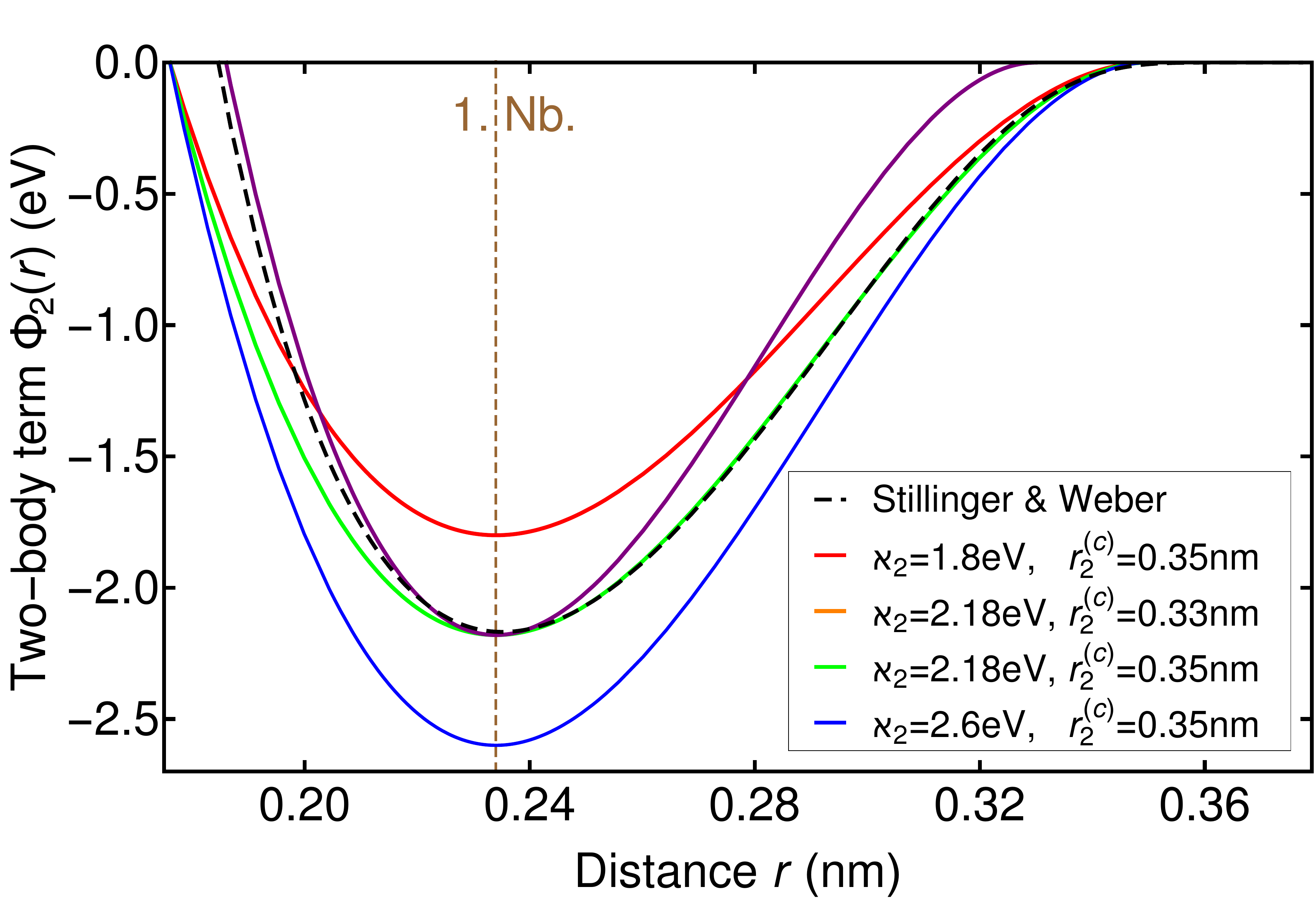}
	\caption{The two-body potential $\Phi^{(\text{SW})}_2$ of Stillinger \& Weber is shown together with the two-body potentials $\Phi^{(\text{pol})}_2$ of selected test potentials. The brown vertical line indicates the distance of the nearest neighbors in Si.}
	\label{fig:phi2_test_pots}
\end{figure}

As a starting point of our study, we set the strength $\aleph_2=2.18$ eV and the cutoff-radius $r^{(\text{c})}_2=0.35$ nm to get a similar course of the polynomial two-body potential $\Phi^{(\text{pol})}_2$ compared with the Stillinger \& Weber two-body potential $\Phi^{(\text{SW})}_2$ for distances bigger than the first neighbor distance $0.234$ nm, as one can see in Fig. \ref{fig:phi2_test_pots}.

\begin{figure}[htb!]
	\centering
	\includegraphics[width=0.48\textwidth]{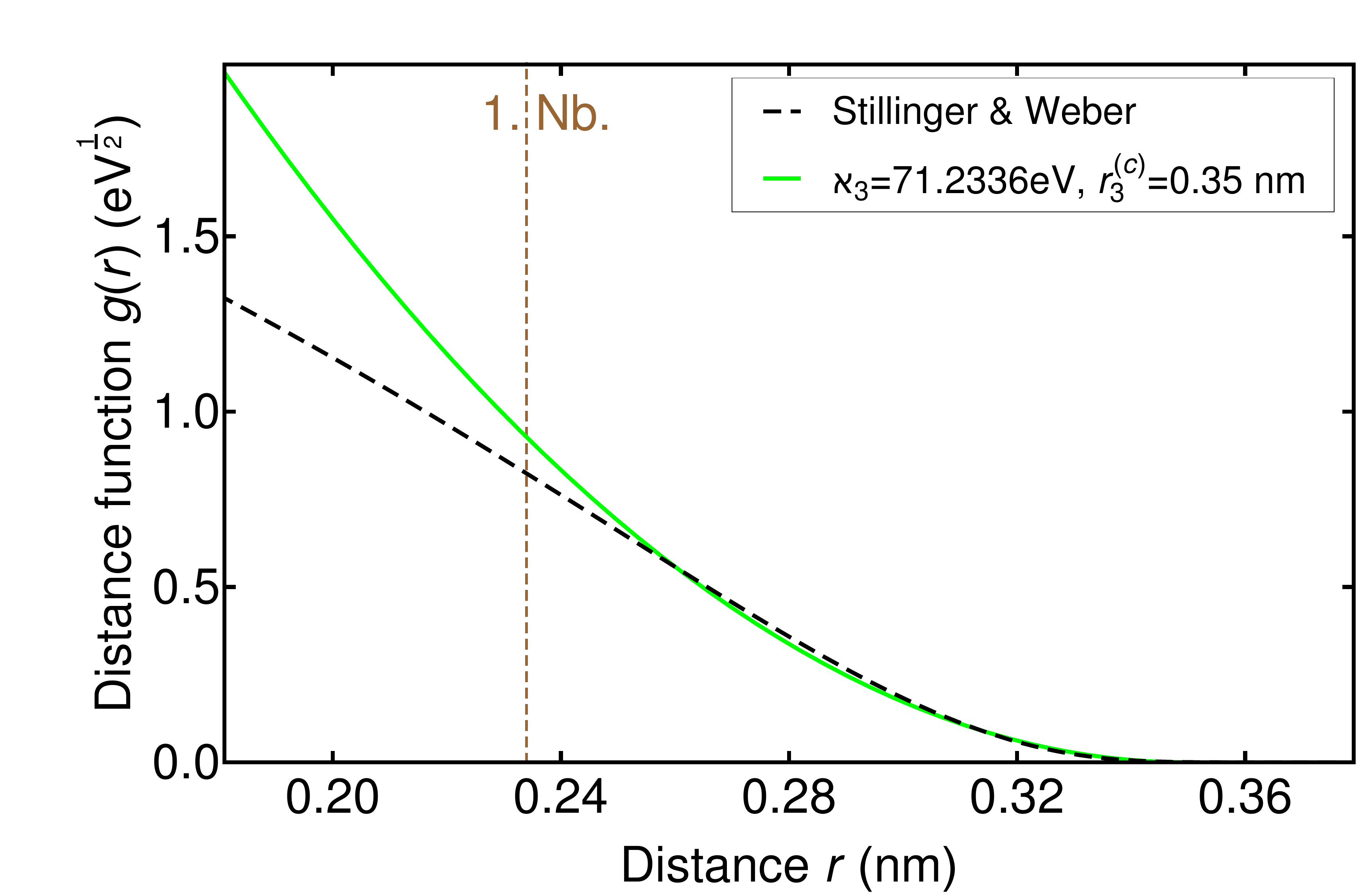}
	\caption{Distance function $g(r)$ of the Stillinger \& Weber and the test potential with $\Phi^{(\text{SW})}_2 \approx \Phi^{(\text{pol})}_2$ and same $T_\text{m}$ is shown. The brown vertical line indicates the distance of the nearest neighbors in Si.}
	\label{fig:distance_func_test_pots}
\end{figure}

We varied the strength $\aleph_3$ of the corresponding three-body potential $\Phi^{(\text{pol})}_3$ in order to get the same melting temperature at zero pressure like the Stillinger \& Weber potential.
In this way we found $\aleph_3=71.2336$ eV and, for this value, $g^{(\text{pol})}(r)$ is similar to $g^{(\text{SW})}(r)$ for distances bigger than $0.25$ nm, as one can see in FIG. \ref{fig:distance_func_test_pots}.
To determine $\aleph_3$, we initially performed several small cell liquid-crystal coexistence MD simulations to get a prediction of the corresponding $\aleph_3$ value.
Then we performed large scale liquid-crystal coexistence MD simulations for a few $\aleph_3$ values to get the searched value of $\aleph_3=71.2336$ eV.
We repeated this whole procedure for several strengths $\aleph_2$ of the polynomial two-body potential to get the corresponding $\aleph_3$ values for reaching the same melting temperature at zero pressure.
We also reduced the cutoff-radius of the polynomial two-body potential to $r^{(\text{c})}_2=0.33$ nm at $\aleph_2=2.18$ eV and determined the corresponding $\aleph_3$.
We list the obtained results in Tab. \ref{tab:Tm_testpots}.
In addition, we increased the cutoff-radius of the polynomial two-body potential to $r^{(\text{c})}_2=0.37$ nm at $\aleph_2=2.18$ eV.
But this setting leads to a crystallization in the hexagonal closed-packed (hcp) structure instead of the diamond-like structure.
Consequently, we skipped this parameter combination in Tab. \ref{tab:Tm_testpots}.

\begin{table*}[htb!]
	\caption{\label{tab:Tm_testpots} 
		Melting temperature $\bigl.T_\text{m}\bigr|_{p=0}$ and slope $\bigl.\frac{dT_\text{m}}{dp}\bigr|_{p=0}$ at zero pressure are listed for the different test and the Stillinger \& Weber potential.}
	\centering
	\begin{tabular}{llllllcr}
		\toprule
		$\aleph_2$ (eV) && $\aleph_3$ (eV) && $r^{(\text{c})}_2$ (nm) & $r^{(\text{c})}_3$ (nm) & $\bigl.T_\text{m}\bigr|_{p=0}$ (K) & $\left.\frac{dT_\text{m}}{dp}\right|_{p=0}$ $\left(\frac{\text{K}}{\text{GPa}}\right)$ \\
		\hline
		\multicolumn{4}{l}{Stillinger \& Weber} & 0.377118 & 0.377118 & $1688 \pm 3$ & $-49 \pm 3$ \\[6pt]
		1.8   && 61.999876 && 0.35     & 0.35     & $1689 \pm 3$ & $-29 \pm 4$ \\
		2.0   && 66.7489   && 0.35     & 0.35     & $1688 \pm 3$ & $-43 \pm 3$ \\
		2.18  && 71.2336   && 0.35     & 0.35     & $1689 \pm 3$ & $-56 \pm 4$ \\
		2.2   && 71.723961 && 0.35     & 0.35     & $1687 \pm 3$ & $-58 \pm 4$ \\
		2.4   && 76.9831   && 0.35     & 0.35     & $1689 \pm 3$ & $-68 \pm 3$ \\
		2.6   && 82.337476 && 0.35     & 0.35     & $1689 \pm 3$ & $-89 \pm 3$ \\[6pt]
		2.18  && 45.104656 && 0.33     & 0.35     & $1688 \pm 3$ & $-80 \pm 6$ \\
		\hline
	\end{tabular}
\end{table*}

\begin{figure}[htb!]
	\centering
	\includegraphics[width=0.48\textwidth]{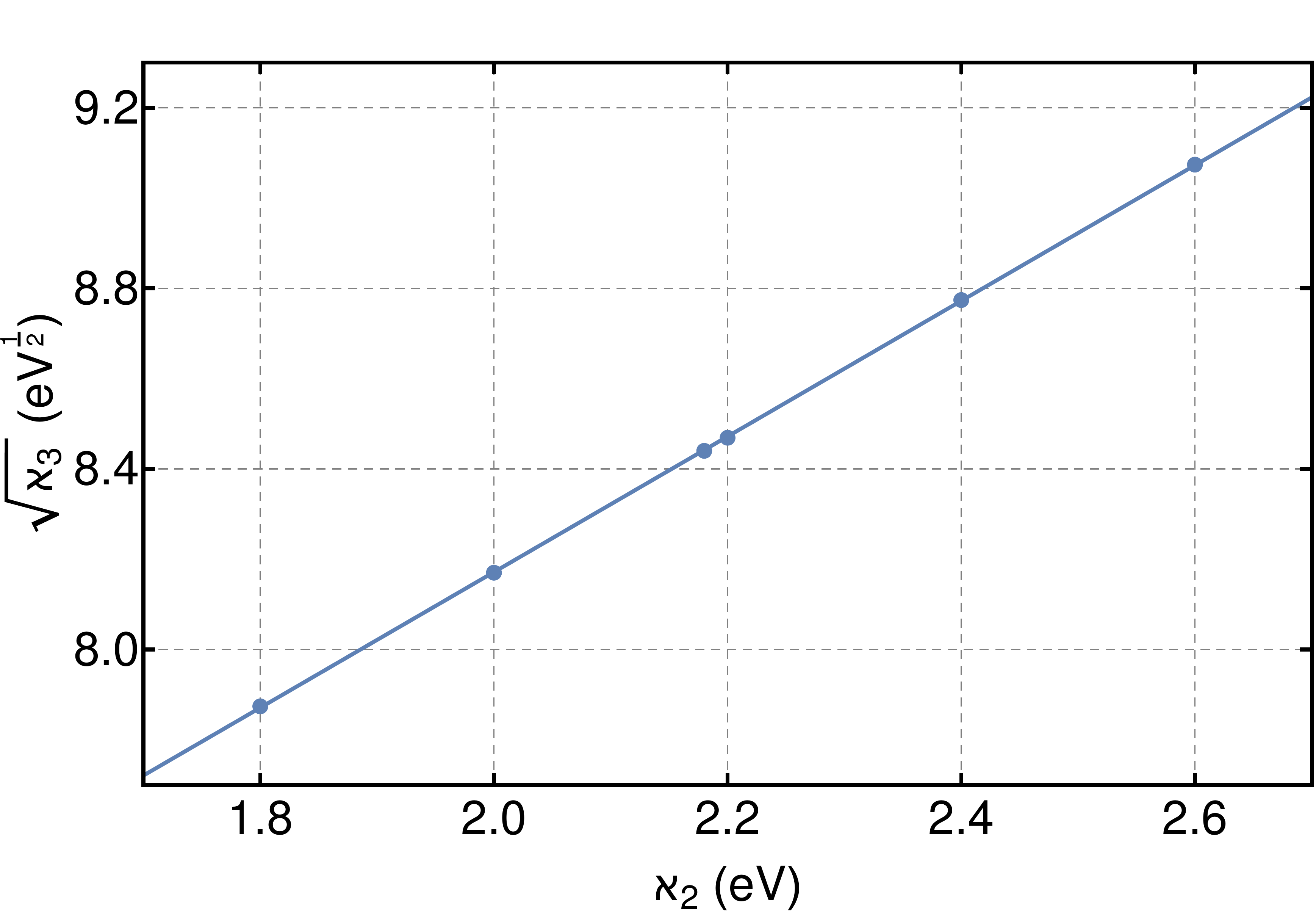}
	\caption{$\sqrt{\aleph_3}$ is shown as a function of $\aleph_2$ for the test potentials with $r^{(\text{c})}_2=r^{(\text{c})}_3=0.35$ nm and $T_\text{m}\approx 1687$ K (see TAB. \ref{tab:Tm_testpots}). 
		Also a linear regression line is inserted.}
	\label{fig:aleph2_sqrt_aleph3}
\end{figure}

Our study shows that the strength $\aleph_2$ of the two-body potential is not relevant for the melting temperature $\bigl.T_\text{m}\bigr|_{p=0}$ at zero pressure, as one can be seen in Tab. \ref{tab:Tm_testpots}.
Rather, $\bigl.T_\text{m}\bigr|_{p=0}$ is defined by the relationship of $\aleph_2$ and $\aleph_3$.
Fig. \ref{fig:aleph2_sqrt_aleph3} shows $\sqrt{\aleph_3}$ as a function of $\aleph_2$ for the test potentials with $r^{(\text{c})}_2=r^{(\text{c})}_3=0.35$ nm and $\bigl.T_\text{m}\bigr|_{p=0}\approx 1687$ K. 
All these values lie on a straight line, which can be obtained from a linear regression.
In order to reach the same melting temperature, $\aleph_3$ has to increase quadratically for increasing $\aleph_2$ and $\aleph_3$ has to decrease quadratically for decreasing $\aleph_2$.
It is obvious that this dependence is only valid in a certain interval for $\aleph_2$, since the nearest neighbor distance is not stabilized any more for $\aleph_2\to 0$ and the crystal will melt easily at very low temperatures.

The two-body potential is responsible that the nearest neighbors are located at the distance, which corresponds to its minimum.
If no three-body potential is present, each atom prefers as much neighbors as possible at this distance, which is fulfilled for a closed-packed structure like fcc or hcp.
But the presence of the three-body potential forces the nearest neighbors to exhibit always an angle $\theta$ obeying $\cos(\theta)=-\frac{1}{3}$.
This condition is only possible for four nearest neighbors building a tetragonal structure. 
Consequently, the diamond-like structure is formed, which is called open, because free space exists between the atoms, to which the atoms could move during melting.
In addition, if the cutoff radius of the two-body potential is bigger than that one of the three-body potential, the atoms still crystallize in the hcp structure like it would be without three-body potential.

An increase of the pressure $p$ induces a closer location of the atoms to each other. 
Now, the nearest neighbors are located to close to each other and the two-body potential associated forces move them further away. 
Consequently, the nearest neighbors can move more easily into the free space and the crystal will melt more easily, {\it i.e.}, $T_\text{m}$ decreases with increasing pressure.
On the other hand, a decrease of the pressure induces bigger distances between the atoms.
Then, the nearest neighbors are located to far away to each other and the two-body potential associated forces move them closer together.
Consequently, the nearest neighbors can move less easily into the free space and the crystal will melt more hardly, {\it i.e.}, $T_\text{m}$ increases with decreasing pressure.
This is the explanation of the negative slope.

Moreover, a bigger increase besides the minimum of the two-body potential, like for increasing $\aleph_2$ or decreasing $r^{(\text{c})}_2$ (see Tab. \ref{tab:Tm_testpots}), induces a more negative slope, because the pressure-conditioned displacement of the nearest neighbors from their equilibrium distance will cause stronger forces on them.
Hence, we should add a function with a minimum at the nearest neighbor distance and a strong increase beside this minimum to the two-body potential of the polynomial Si potential in order to achieve a negative slope.

\subsection{Correction of the 2-body and 3-body potential coefficients}

If we use the previous results, we need to modify the two- and three-body potentials of the polynomial Si potential $\Phi^{(\text{Si})}(T_\text{e})$ to control the melting temperature $\bigl.T_\text{m}\bigr|_{p=0}$ and the slope $\bigl.\frac{dT_\text{m}}{dp}\bigr|_{p=0}$ at zero pressure.
Using $\aleph_3$, we performed the manipulation of the three-body potential in exactly the same way as described in Sec. \ref{sec:correction_3body}.
The two-body potential of $\Phi^{(\text{Si})}(T_\text{e})$ contains the cutoff-radius of $r^{(\text{c})}_2=0.63$ nm and has the degree $N^{(r)}_2=10$ (see Sec. \ref{sec:Polypot}).

\begin{figure}[htb!]
	\centering
	\includegraphics[width=0.48\textwidth]{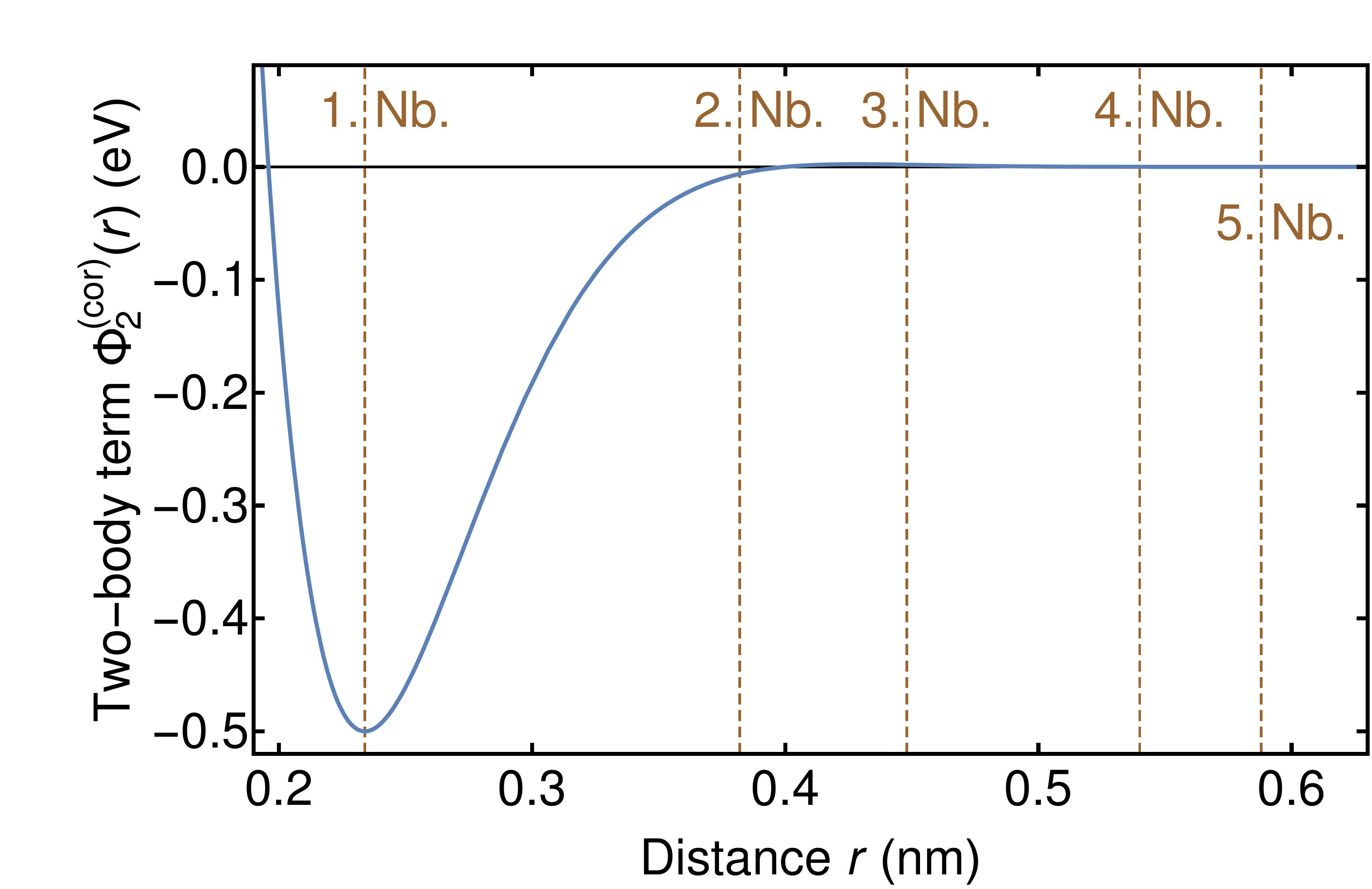}
	\caption{The correction two-body term $\Phi^{(\text{cor})}_2$ is shown for $\aleph_2=0.5$ eV.
		The brown vertical lines indicate the positions of the neighbors in the diamond-like structure of Si.}
	\label{fig:phi2_coeff_corr2}
\end{figure}

In order to get a negative slope, the coefficients $\bigl\{c^{(q)}_2\bigr\}$ of the two-body potential should be modified in the following way:
A correction two-body term $\Phi^{(\text{cor})}_2$ should be added, which exhibits a single minimum at the nearest neighbor distance 0.234 nm of the diamond-like structure of Si and which increases strongly beside the minimum.
To do do, we used a linear combination of the three highest powers of the term $\bigl(1-r_{ij} / r^{(\text{c})}_2\bigr)$ for $\Phi^{(\text{cor})}_2$.
The three highest powers were chosen for the manipulation, since lower powers induce a weaker increase beside the minimum of $\Phi^{(\text{cor})}_2$.
We derive the coefficients of $\Phi^{(\text{cor})}_2$ from the following conditions: 
$\Phi^{(\text{cor})}_2$ exhibits a minimum at $r^{(\text{min})}=0.234$ nm and sets to -$\aleph_2$ at this minimum and sets to zero at $r^{(1)}=0.4$ nm.
We introduce the last constraint, since $\Phi^{(\text{cor})}_2$ should stay approximately at 0 for distances between $r^{(1)}$ and the cutoff radius $r^{(\text{c})}_2=0.63$ nm. 
$r^{(1)}$ should be as small as possible, because the increase beside the minimum becomes bigger for decreasing $r^{(1)}$, which allows a stronger slope $\bigl.\frac{dT_\text{m}}{dp}\bigr|_{p=0}$ control.
But, if $r^{(1)}$ becomes smaller than $0.4$ nm, $\Phi^{(\text{cor})}_2$ becomes significantly positive for distances bigger than $r^{(1)}$, which should be avoided.
We present the correction two-body term $\Phi^{(\text{cor})}_2$, which fulfills the above mentioned conditions, in Fig. \ref{fig:phi2_coeff_corr2}.
It is constructed as
\begin{eqnarray}
\Phi^{(\text{cor})}_2(r_{ij}) &=& \hspace{9pt} 647.3458562015724 \ \aleph_2  \left(1-\frac{r_{ij}}{r^{(\text{c})}_2}\right)^8 \nonumber \\
&&- 2712.579093531517 \ \aleph_2  \left(1-\frac{r_{ij}}{r^{(\text{c})}_2}\right)^9  \nonumber \\
&& + 2573.178456073968 \ \aleph_2  \left(1-\frac{r_{ij}}{r^{(\text{c})}_2}\right)^{10}.
\label{equ:Phi2_coeff_corr2}
\end{eqnarray}

Adding the above mentioned correction two-body term $\Phi^{(\text{cor})}_2$ to $\Phi_2$ corresponds just to modify the coefficients $c^{(2)}_8$, $c^{(2)}_9$ and $c^{(2)}_{10}$ of $\Phi_2$.
More detailed, $647.3458562015724 \, \aleph_2$ is add to $c^{(8)}_2$, $- 2712.579093531517 \, \aleph_2$ to $c^{(9)}_2$ and $2573.178456073968 \, \aleph_2$ to $c^{(10)}_2$.

\begin{figure*}[htb!]
	\centering
	\includegraphics[width=0.48\textwidth]{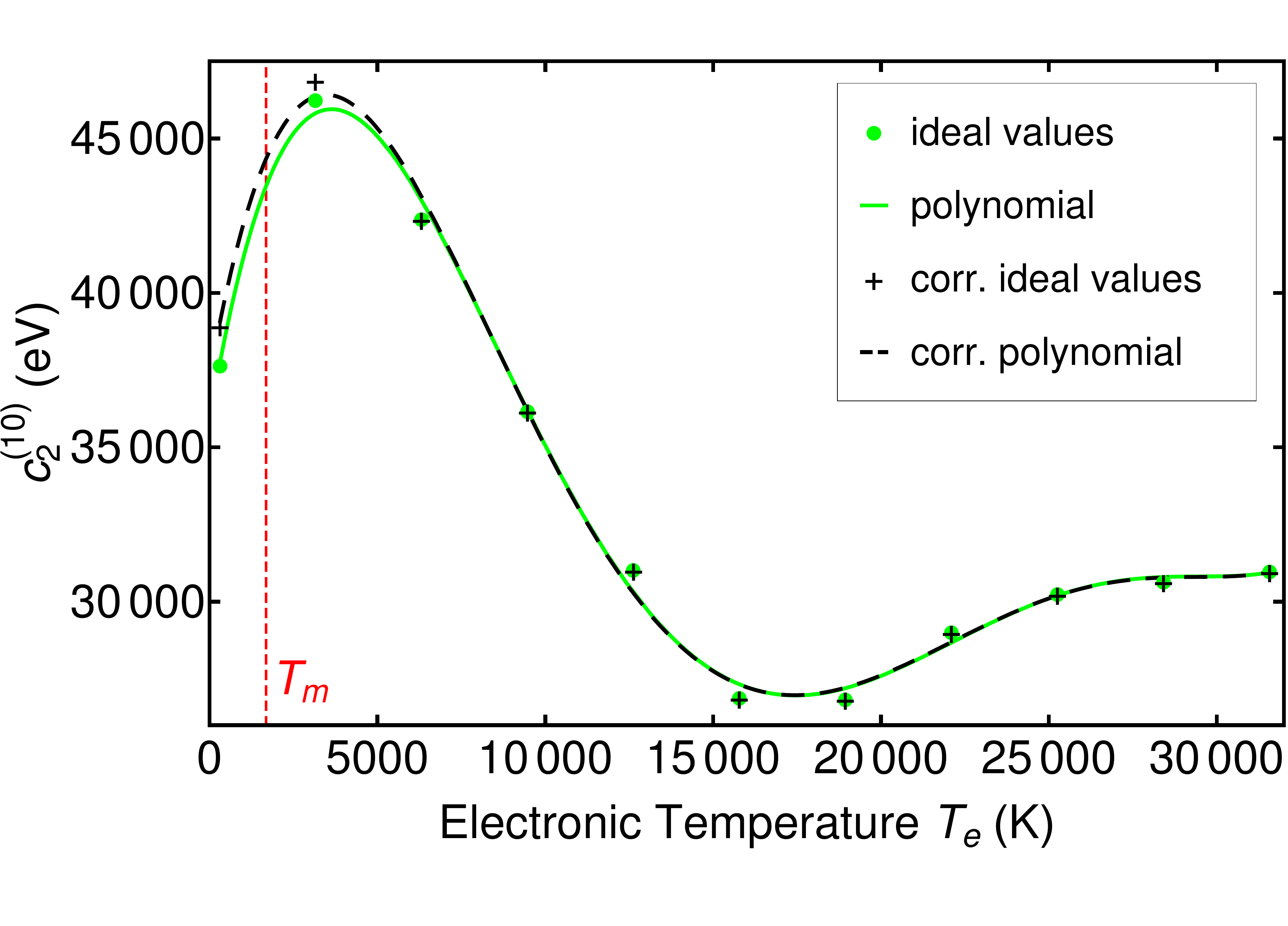}
	\includegraphics[width=0.48\textwidth]{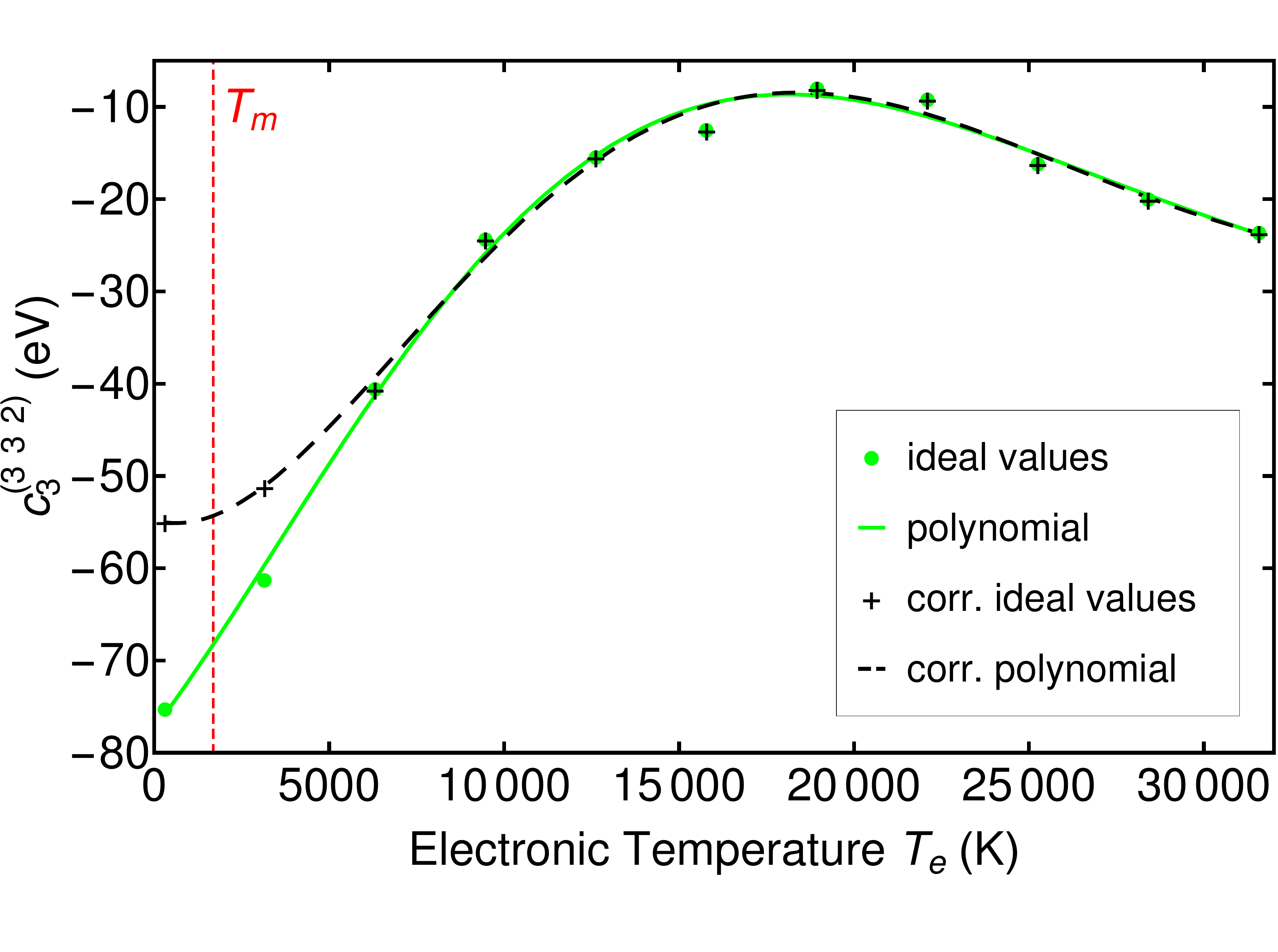}
	\includegraphics[width=0.48\textwidth]{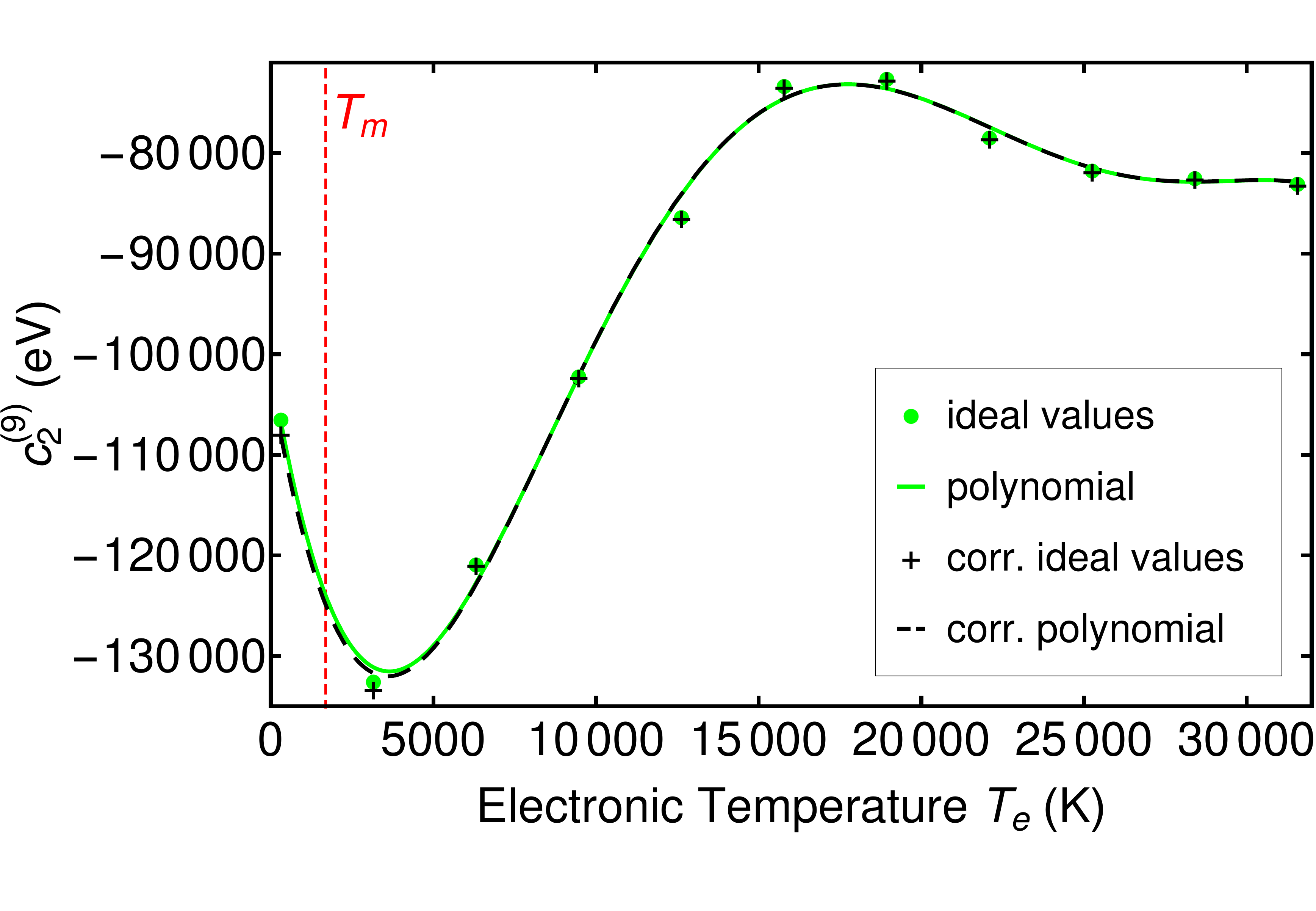}
	\includegraphics[width=0.48\textwidth]{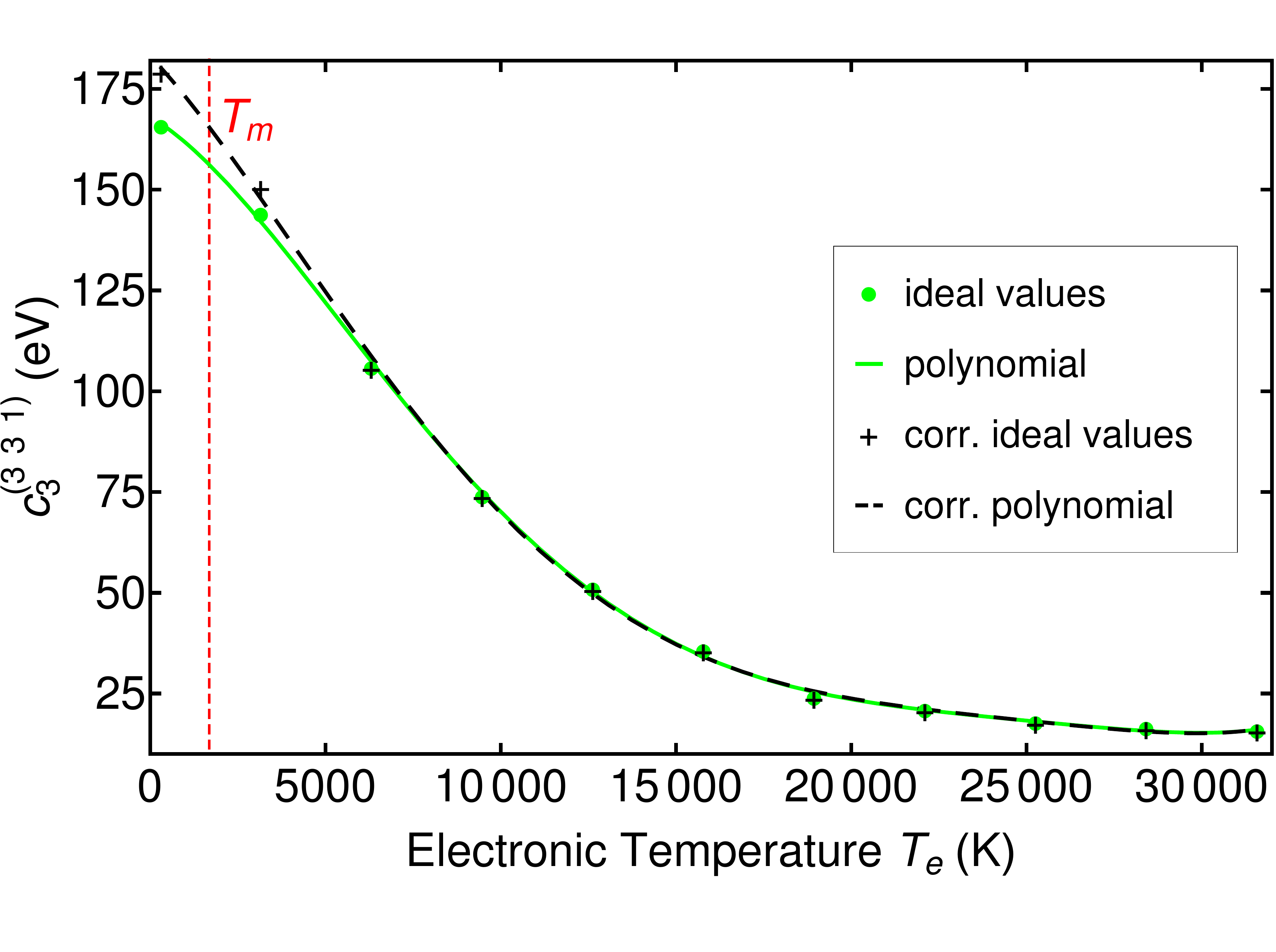}
	\includegraphics[width=0.48\textwidth]{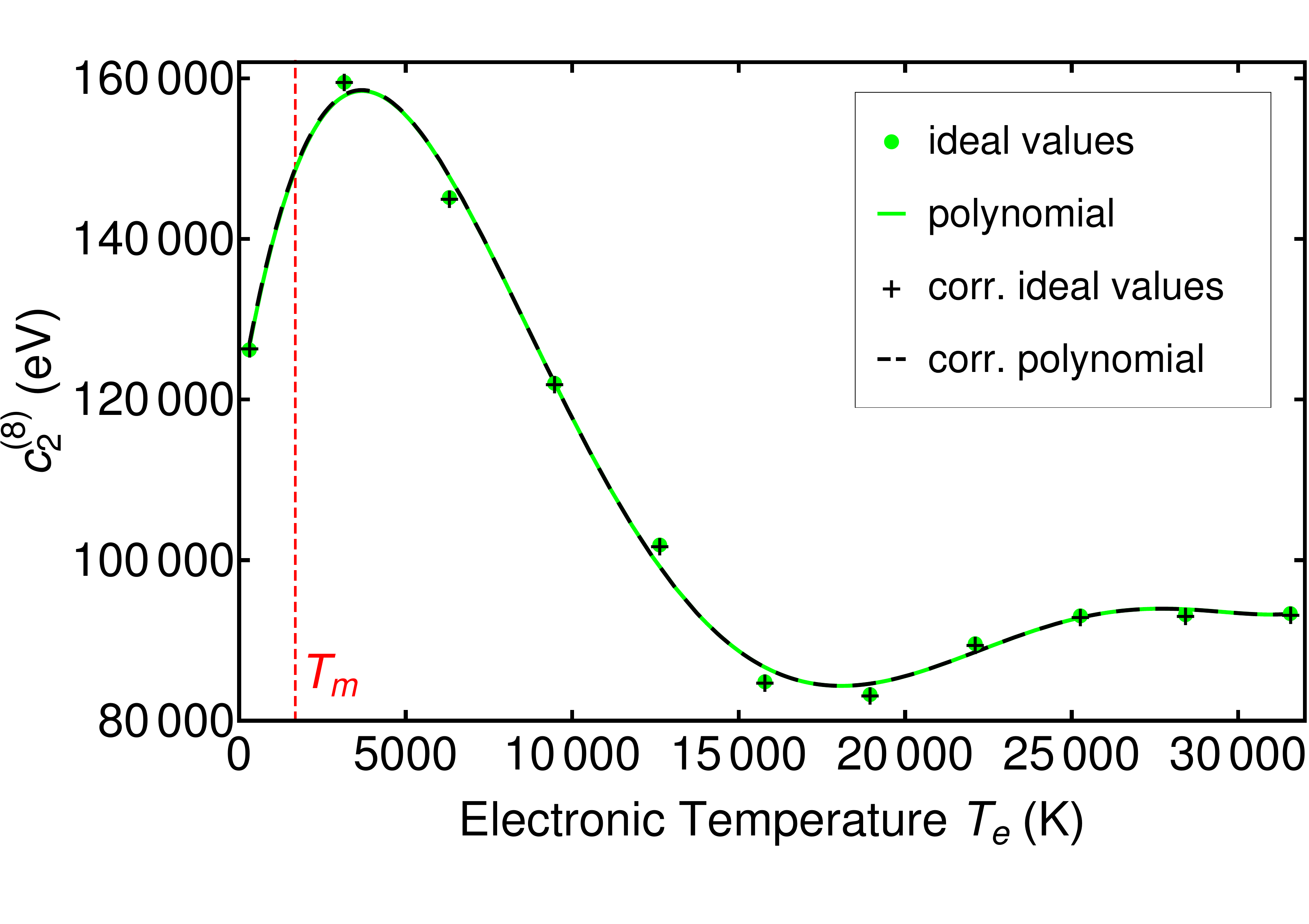}
	\includegraphics[width=0.48\textwidth]{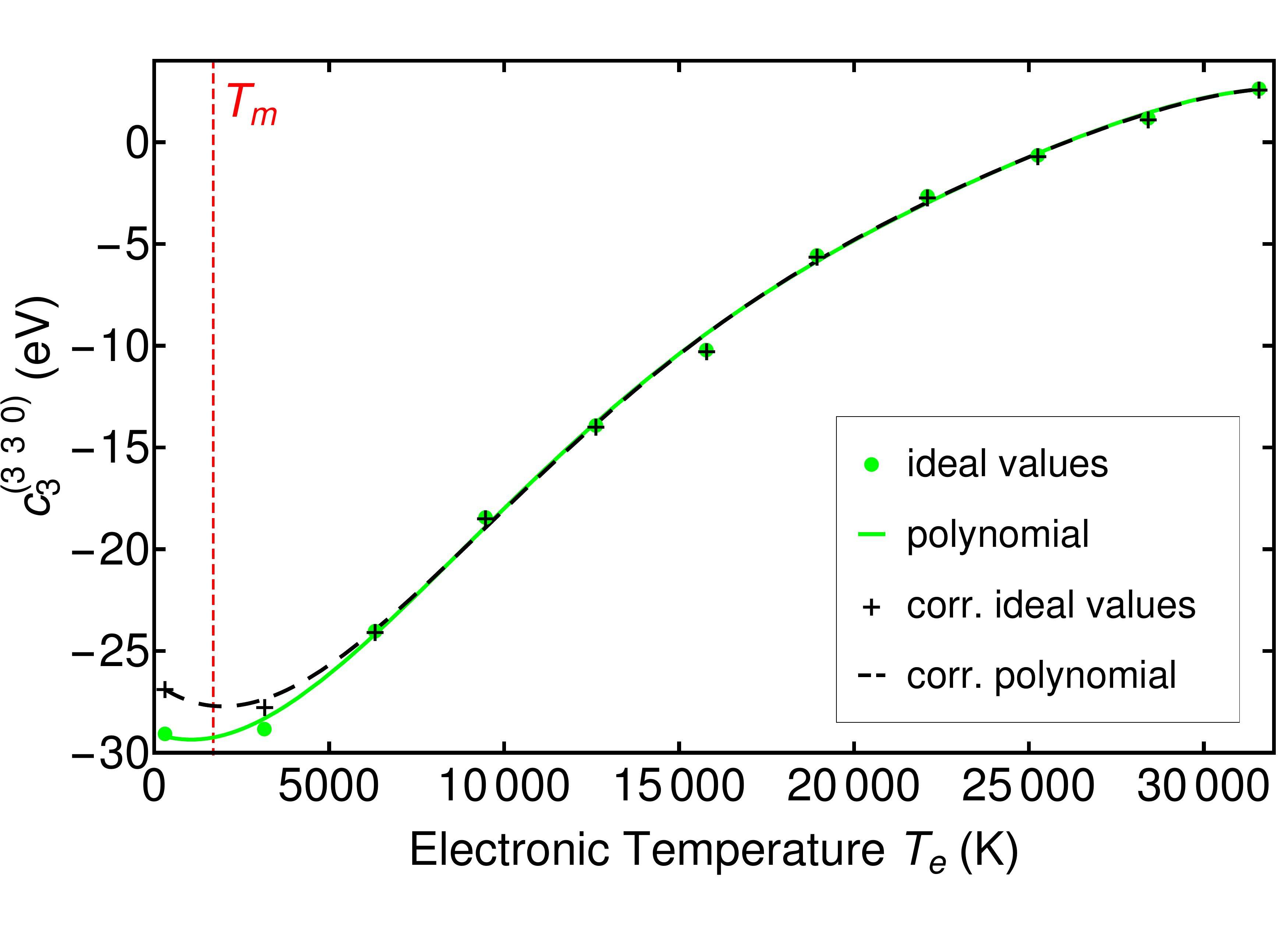}
	\caption{The polynomial approximation of the coefficients $c^{(2)}_{8}$, $c^{(2)}_{9}$, $c^{(2)}_{10}$, $c^{(3)}_{3\,3\,0}$, $c^{(3)}_{3\,3\,1}$, and $c^{(3)}_{3\,3\,2}$ is shown together with the ideal values before (green) and after (black) the correction of the ideal values at $T_\text{e}= 316$ K and $T_\text{e}= 3158$ K. $\aleph_2=0.5$ eV and $\aleph_3=20.3$ eV are shown, which yields the experimental melting temperature $T_\text{m}=1687$ K highlighted by a red vertical line.}
	\label{fig:Tm_coeff_corr2}
\end{figure*}

Similar to the $\aleph_3$ modification, we add the corresponding correction value to the ideal coefficient values at $T_\text{e}=316$ K and half of it at $T_\text{e}=3158$ K before the polynomial is approximated from the ideal coefficient values in order to get the smooth $T_\text{e}$-dependence of the potential coefficients.
We show in Fig. \ref{fig:Tm_coeff_corr2} the original and corrected ideal coefficient values together with the corresponding fitted polynomials for the modified coefficients at $\aleph_2=0.5$ eV and $\aleph_3=20.3$ eV.

We gradually increased $\aleph_2$ and, for each $\aleph_2$, we determined the corresponding $\aleph_3$ in such a way that the interatomic potential exhibits the experimental melting temperature of $T_\text{m}=1687$ K at zero pressure. For this, we used the procedure described in Sec. \ref{sec:Tm_slope_study_on_test_potentials}.
In Summary, this study was very computational expensive and needed millions of core hours, which were distributed on three different computer clusters.
We list the finally obtained $\aleph_2$, $\aleph_3$ pairs together with the corresponding melting temperature and slope at zero pressure in Tab. \ref{tab:Tm_correction_coeff}.

\begin{table}[htb!]
	\caption{\label{tab:Tm_correction_coeff}
	Melting temperature $\bigl.T_\text{m}\bigr|_{p=0}$ and slope $\bigl.\frac{dT_\text{m}}{dp}\bigr|_{p=0}$ at zero pressure are listed for the different coefficient corrections to the interatomic potential.}
	\centering
	\begin{tabular}{clclcr}
		\toprule
		$\aleph_2$ (eV) && $\aleph_3$ (eV) && $\bigl.T_\text{m}\bigr|_{p=0}$ (K) & $\left.\frac{dT_\text{m}}{dp}\right|_{p=0}$ $\left(\frac{\text{K}}{\text{GPa}}\right)$ \\
		\hline
		0.0 && 12.0 && $1687 \pm 3$ & $ 18 \pm 4$ \\
		0.1 && 13.1 && $1686 \pm 3$ & $ 12 \pm 4$ \\
		0.2 && 14.6 && $1687 \pm 3$ & $  3 \pm 4$ \\
		0.3 && 16.4 && $1689 \pm 3$ & $ -1 \pm 4$ \\
		0.4 && 18.2 && $1685 \pm 3$ & $ -7 \pm 4$ \\
		0.5 && 20.3 && $1689 \pm 3$ & $-11 \pm 4$ \\
		0.6 && 22.4 && $1688 \pm 3$ & $-14 \pm 4$ \\
		\hline
	\end{tabular}
\end{table}

\begin{figure}[htb!]
	\centering
	\includegraphics[width=0.48\textwidth]{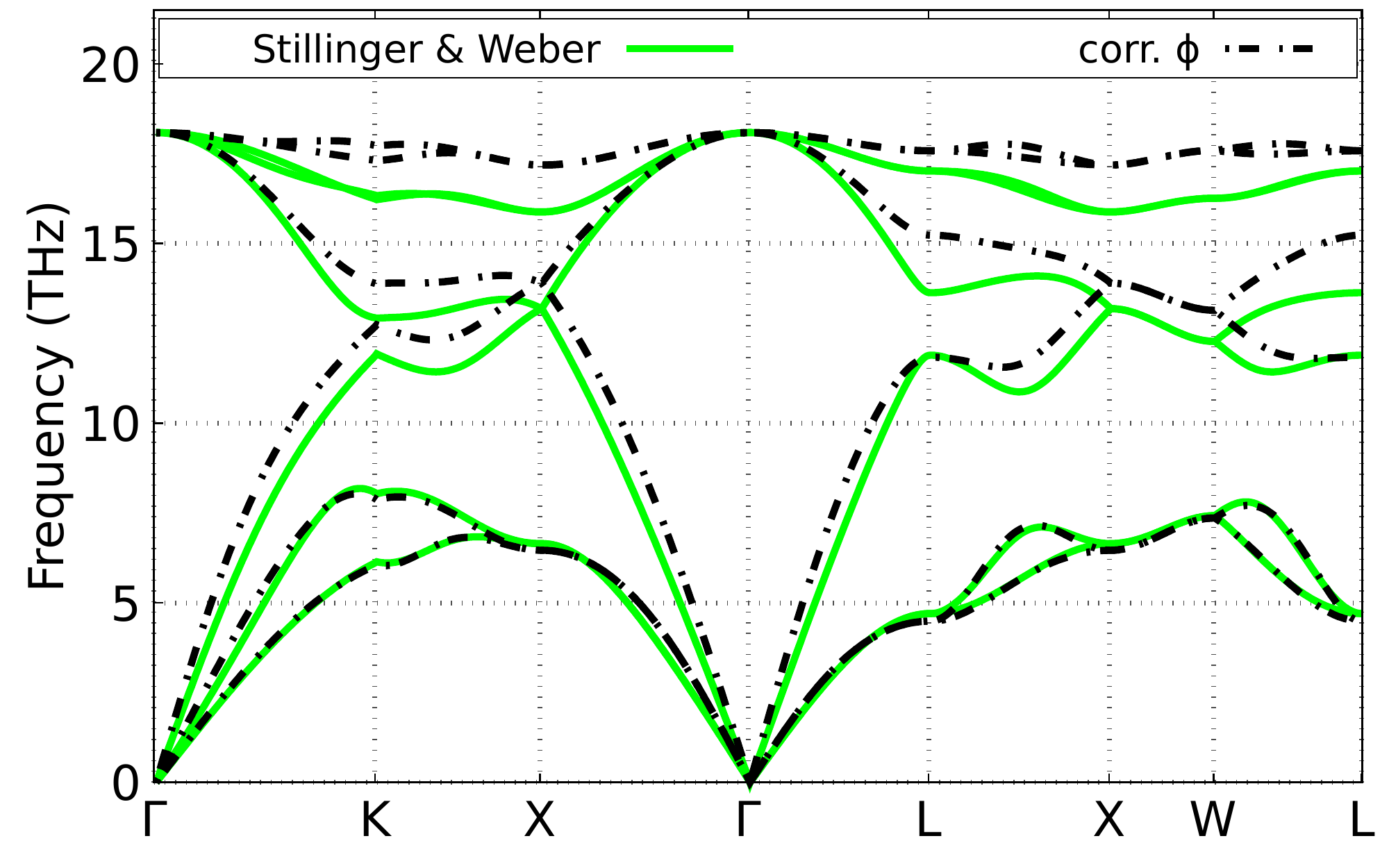}
	\caption{The phonon bandstructure of the diamond-like structure is shown for the $\aleph_2=0.5$ eV, $\aleph_3=20.3$ eV corrected $\Phi$ at $T_\text{e}=316$ K (black dot-dashed) and the Stillinger \& Weber potential (green solid).}
	\label{fig:bandstructure_coeff_corr2_SW}
\end{figure}

One can clearly see in Tab. \ref{tab:Tm_correction_coeff} that the slope decreases with increasing $\aleph_2$ and it becomes negative at $\aleph_2=0.3$ eV.
With increasing $\aleph_2$ value, also the phonon frequencies increase, since the bonding becomes stronger.
At $\aleph_2=0.5$ eV, the corresponding phonon bandstructure of the diamond-like structure at $T_\text{e}=316$ K looks similar to the one of the famous Stillinger \& Weber potential, as one can be seen in Fig. \ref{fig:bandstructure_coeff_corr2_SW}.
Especially the acoustic phonon branches are in an excellent agreement.
Consequently, we selected $\aleph_2=0.5$ eV for the final corrected interatomic potential.
We tabulate the corresponding modified coefficients in the Supplemental Material.
In addition, we provide a Fortran subroutine for the calculation of the cohesive energy and the forces from the final corrected interatomic potential.

\begin{figure}[htb!]
	\centering
	\includegraphics[width=0.48\textwidth]{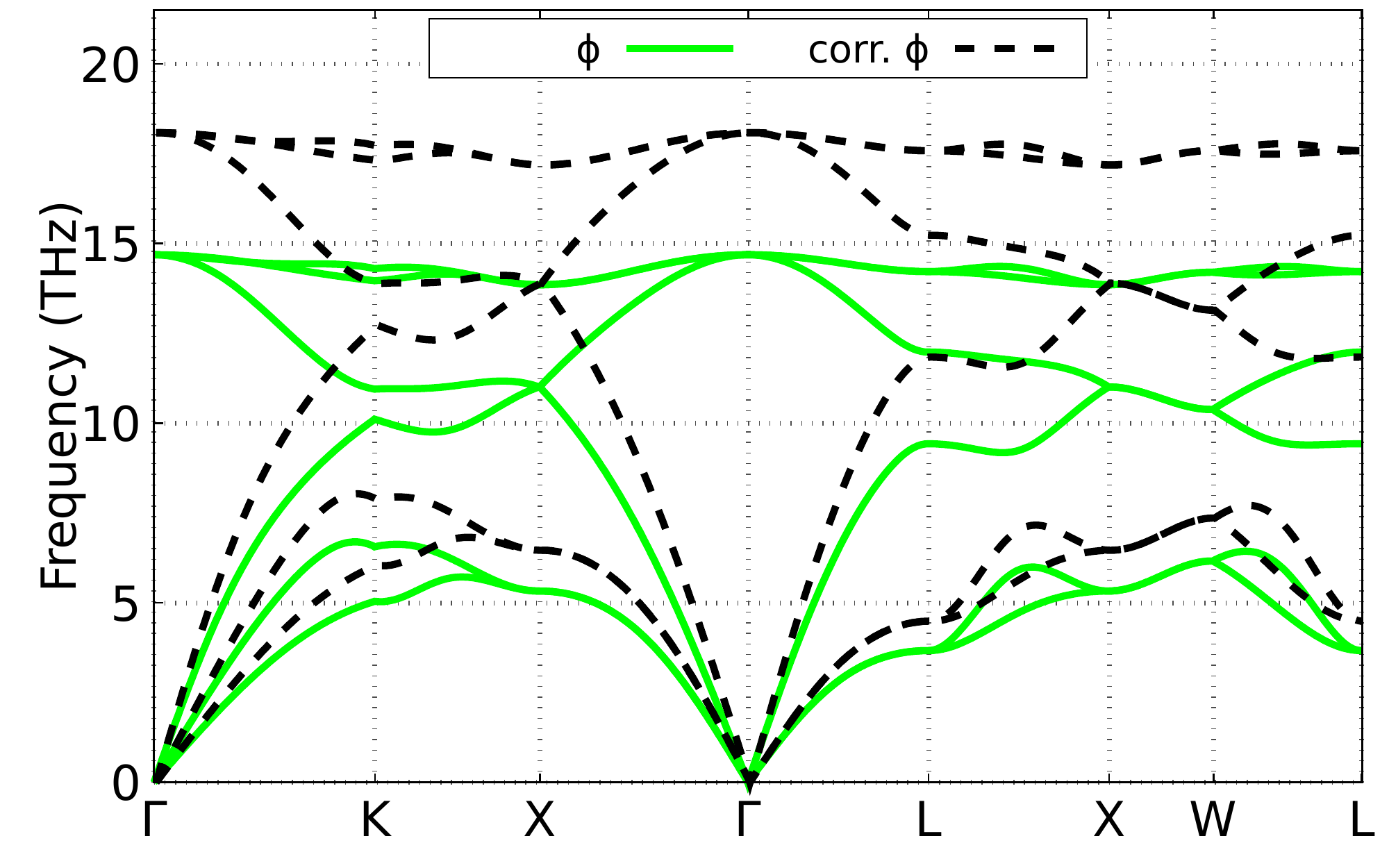}
	\caption{The phonon bandstructure of the diamond-like structure is shown for the $\aleph_2=0.5$ eV, $\aleph_3=20.3$ eV corrected (black dashed) and uncorrected (green solid) interatomic potential $\Phi$ at $T_\text{e}=316$ K.}
	\label{fig:bs_coef_corr2}
\end{figure}

\begin{figure}[htb!]
	\centering
	\includegraphics[width=0.48\textwidth]{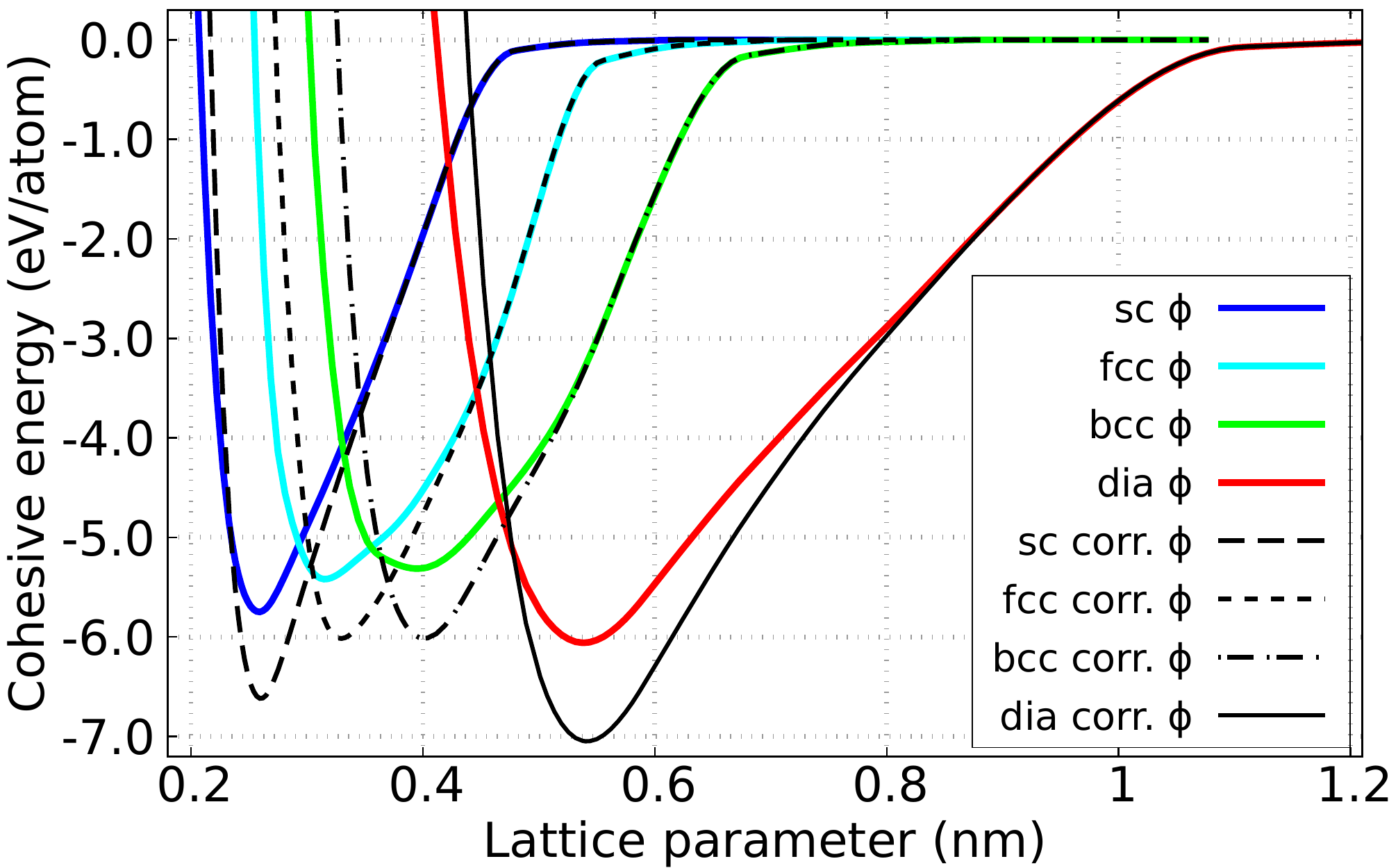}
	\caption{The cohesive energies of different structures are shown for the $\aleph_2=0.5$ eV, $\aleph_3=20.3$ eV corrected (black dashed) and uncorrected (colored solid) interatomic potential $ \Phi$ at $T_\text{e}=316$ K.}
	\label{fig:Ecoh_coef_corr2}
\end{figure}

\begin{figure}[htb!]
	\centering
	\includegraphics[width=0.48\textwidth]{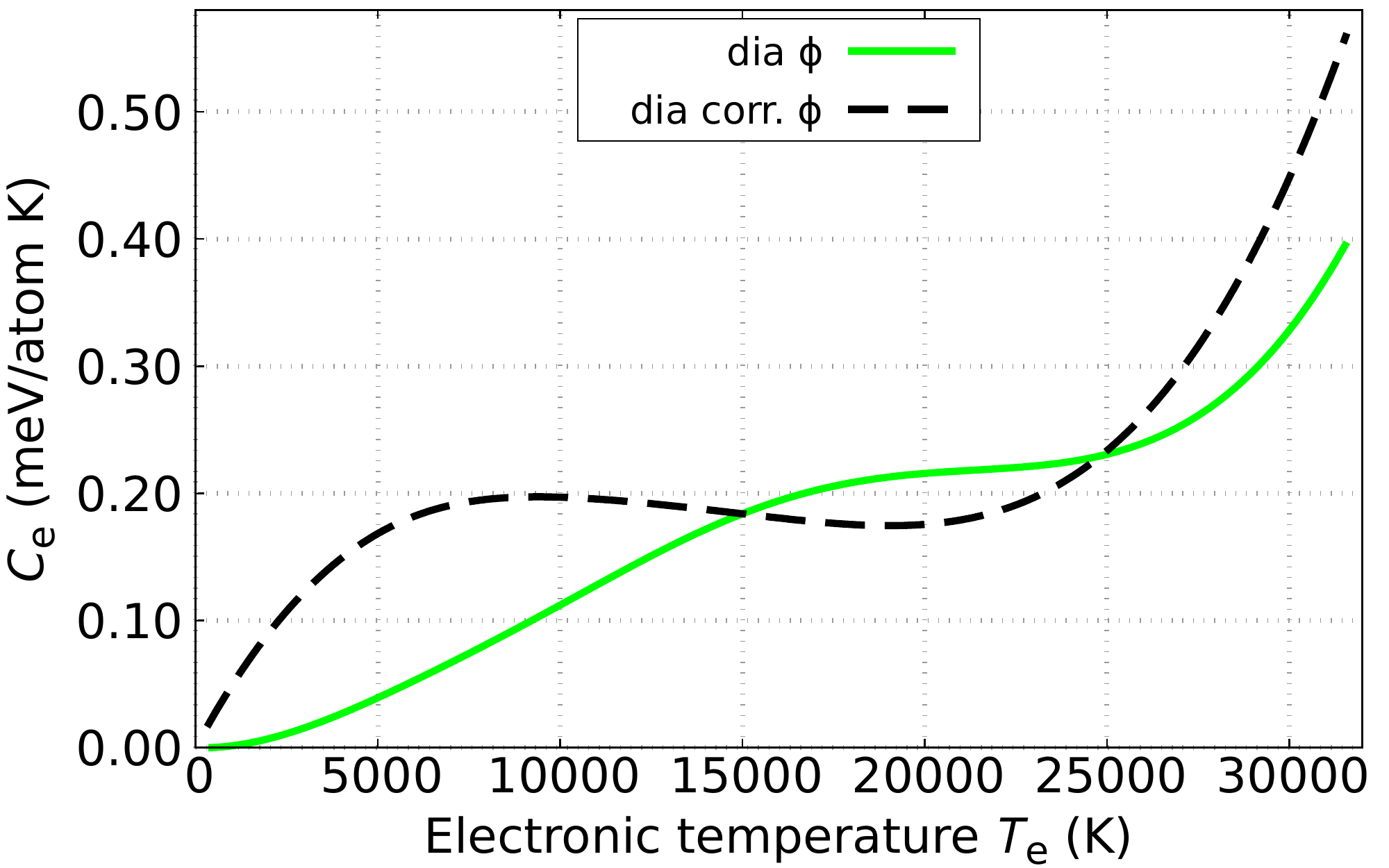}
	\caption{The electronic specific heat $C_\text{e}$ of the diamond-like structure is shown for the $\aleph_2=0.5$ eV, $\aleph_3=20.3$ eV corrected (black dashed) and uncorrected (green solid) interatomic potential $\Phi$ at $T_\text{e}=316$ K.}
	\label{fig:CVe_coef_corr2}
\end{figure}

\begin{figure}[htb!]
	\centering
	\includegraphics[width=0.48\textwidth]{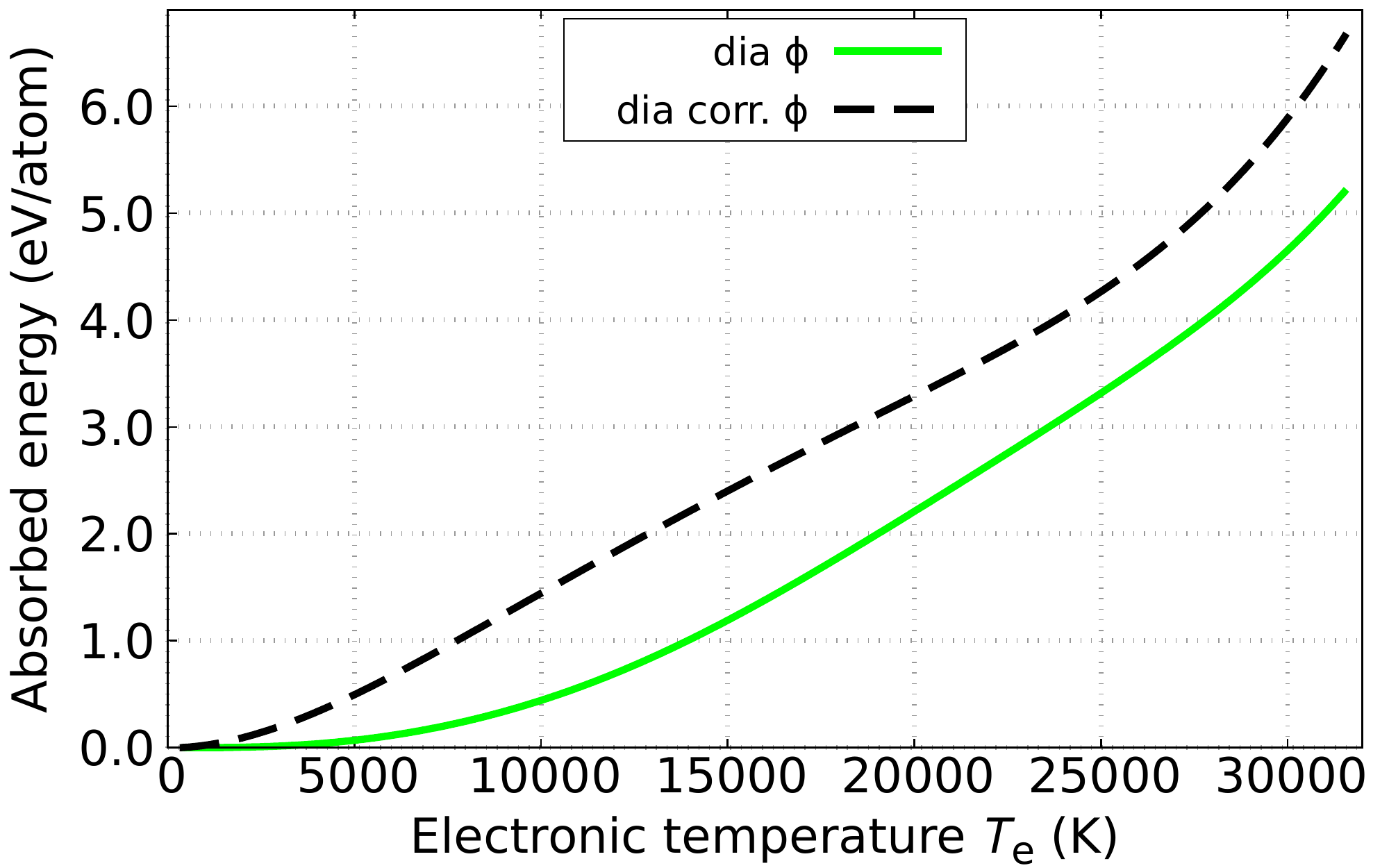}
	\caption{The absorbed electronic energy $E_\text{abs}$ of the diamond-like structure is shown for the $\aleph_2=0.5$ eV, $\aleph_3=20.3$ eV corrected (black dashed) and uncorrected (green solid) interatomic potential $\Phi$ at $T_\text{e}=316$ K.}
	\label{fig:Eabs_coef_corr2}
\end{figure}

Fig. \ref{fig:bs_coef_corr2} presents the comparison of the phonon bandstructure, Fig. \ref{fig:Ecoh_coef_corr2} the comparison of the cohesive energies of several bulk crystal structures, Fig. \ref{fig:Eabs_coef_corr2} the comparison of the absorbed electronic energy, and Fig. \ref{fig:CVe_coef_corr2} the comparison of the electronic specific heat between the corrected and uncorrected interatomic potential at $T_\text{e}=316$ K.
The absorbed energy $U_\text{e}$ and the electronic specific heat are directly calculated from the $T_\text{e}$-dependent interatomic potential from
\begin{eqnarray}
U_\text{e}&=&\Phi -T_\text{e}\,\frac{\partial \Phi}{\partial T_\text{e}} \label{equ:Ue} \\
C_\text{e} &=& -T_\text{e} \frac{\partial^2 \Phi}{\partial T^2_\text{e}}.
\label{equ:Ce}
\end{eqnarray}

The corrected potential contains a stronger two-body term.
Thus, the phonon frequencies and the absolute value of the cohesive energies increase after the modification.
The $T_\text{e}$-dependence of the electronic specific heat is significantly changed by the modification.
We can still accept it, since it is positive.
The absorbed electronic energy of the corrected interatomic potential is $\sim 1.0$ eV higher compared to the uncorrected one for $T_\text{e}>10000$ K, but the functional shape is the same.

Among the data used for fitting, we included ab-initio structural free cohesive energies and ab-initio atomic forces from atomic configurations of molecular dynamics simulations at constant $T_\text{e}$.
We also calculated the relative error in the atomic forces and the structural free cohesive energies of the atomic configurations from these molecular dynamics simulations for the corrected interatomic potential and compared it with the uncorrected one in Tab. \ref{tab:Errors_coeff_correction2}.
The correction does not induce any significant changes in the relative errors at high electronic temperatures $T_\text{e}\geq 9473$ K.
This is not surprising, because the corrected polynomial does not differ from the uncorrected polynomial for the smooth $T_\text{e}$-approximation of the coefficients at high $T_\text{e}$'s, as it can be seen in Fig. \ref{fig:Tm_coeff_corr2}.
Consequently, the melting temperature correction does not influence the physical properties at high $T_\text{e}$.

\begin{table}[htb!]
	\caption{Relative error in the atomic forces $f_\text{err}$ and in the structural free cohesive energies $E_\text{err}$ of the molecular dynamics simulation are listed for the uncorrected and the $\aleph=0.5$ eV, $\aleph_3=20.3$ eV corrected interatomic potential.}
	\label{tab:Errors_coeff_correction2}
	\centering
	\begin{tabular}{rrrrrrr}
		\toprule
		&& & && \multicolumn{2}{c}{$\aleph_2=0.5$ eV $\, \,$} \\
		&& \multicolumn{2}{c}{uncorrected} && \multicolumn{2}{c}{$\aleph_3=20.3$ eV} \\
		$T_\text{e}$ (K) && $f_\text{err}$ (\%) & $E_\text{err}$ (\%) && $f_\text{err}$ (\%) & $E_\text{err}$ (\%)\\
		\hline
		316 && 25.8 & 1.1 && 65.7 & 14.6\\
		3158 && 20.5 & 0.7 && 32.0 &  6.9\\
		6315 && 13.9 & 0.4 && 14.7 &  1.7\\
		9473 &&  9.8 & 0.3 && 10.0 &  0.9\\
		12631 &&  7.7 & 0.5 &&  8.2 &  1.6\\
		15789 &&  7.3 & 0.8 &&  7.5 &  1.1\\
		18946 && 11.2 & 0.6 && 11.2 &  0.2\\
		22104 &&  8.9 & 0.2 &&  9.0 &  1.0\\
		25262 &&  6.6 & 0.6 &&  6.6 &  0.6\\
		28420 &&  6.3 & 1.6 &&  6.4 &  4.3\\
		31577 &&  6.6 & 1.1 &&  6.6 &  2.9\\
		\hline
	\end{tabular}
\end{table}

\section{Conclusions}

We presented a modification of the coefficients of the two-body and three-body potential of our $T_\text{e}$-dependent interatomic potential for Si of Ref. \cite{Bauerhenne2020} that increase the melting temperature to the experimental value of $T_\text{m}=1687$ K \cite{Jayaraman1963} while maintaining a negative slope in the melting temperature vs. pressure diagram.
The modification of only the coefficients of the three-body potential allows to increase the melting temperature to the experimental value but induces an unphysical positive slope.
Thus, also the coefficients of the two-body potential must be modified, which induces finally a negative slope.
The final corrected interatomic potential exhibits still a physically meaningful electronic specific heat and the physical properties at high $T_\text{e}$'s are not influenced.

We want to point out that such a modification of the coefficients is impossible for the commonly used machine learning potentials like neural networks.
Only the construction of the interatomic potential as a sum of physically interpretable terms together with the simple functional form of these terms allow such an adjustment of features that cannot be directly fitted.

\begin{acknowledgments}
This work was supported by the DFG through the grant GA 465/15-2.
B.B. acknowledges the support by the "Promotionsstipendium des Otto-Braun Fonds" and by the "Abschlussstipendium der Universit\"at Kassel".
Computations were performed on the Lichtenberg High Performance Computer (HHLR) TU Darmstadt, on the IT Servicecenter (ITS) University of Kassel, and on the computing cluster FUCHS University of Frankfurt.
\end{acknowledgments}

\end{document}